\documentclass{jfm}
\usepackage[pdfstartview=FitH,bookmarksopenlevel=1]{hyperref}       

\input tcilatex

\begin{document}

\title[Paradoxical predictions of liquid curtains with surface tension]{Paradoxical predictions\\of liquid curtains with surface tension}
\author[E. S. Benilov]{E. S. Benilov\footnote{Email address for correspondence: Eugene.Benilov@ul.ie}}
\affiliation{Department of Mathematics and Statistics, University of Limerick,\\Limerick V94 T9PX, Ireland}
\maketitle
\date{}

\begin{abstract}
This paper examines two-dimensional liquid curtains ejected at an angle to
the horizontal and affected by gravity and surface tension. The flow is, to
leading order, shearless and viscosity, negligible. The Froude number is
large, so that the radius of the curtain's curvature exceeds its thickness.
The Weber number is close to unity, so that the forces of inertia and
surface tension are almost perfectly balanced. An asymptotic equation is
derived under these assumptions, and its steady solutions are explored. It
is shown that, for a given pair of ejection velocity/angle, infinitely many
solutions exist, each representing a steady curtain with a stationary
capillary wave superposed on it. These solutions describe a rich variety of
behaviours: in addition to arching downwards, curtains can \emph{zigzag}
downwards, \emph{self-intersect}, and even \emph{rise }until the initial
supply of the liquid's kinetic energy is used up. The last type of solutions
corresponds to a separatrix between upward- and downward-bending curtains --
in both cases, self-intersecting (such solutions are meaningful only until
the first intersection, after which the liquid just splashes down). Finally,
suggestions are made as to how the existence of upward-bending curtains can
be tested experimentally.
\end{abstract}

\section{Introduction}

A vertical liquid curtain can be created by cutting a long slot of constant
width in the bottom of a tank; once the tank is filled, a flat liquid sheet
will be squeezed through the slot. If the tank is tipped, the liquid will be
squeezed obliquely, and the curtain's trajectory will be curved due to
gravity. Liquid curtains have important industrial applications (e.g.,
manufacturing of paper), but they are also part of classical fluid mechanics
and, as such, have been studied for more than sixty years.

The present paper looks into a highly counter-intuitive phenomenon, so far
examined only theoretically: that of oblique curtains bending upwards, i.e.,
\emph{against gravity}.

The first such example was produced by \cite{KellerWeitz57} using a set of
equations for a slender oblique curtain without shear and viscosity, but
affected by gravity and surface tension. It was shown that all curtains with
the ejection velocity below a certain threshold would bend upwards
regardless of the ejection angle. The threshold corresponds to the Weber
number $We$ being equal to unity ($We$ is the ratio of forces of inertia and
surface tension). Keller \& Weitz interpreted their paradoxical result using
an analogy with a body affected by gravity and an extra force proportional
to the acceleration.

Unfortunately, this thought-provoking work has been virtually forgotten:
since 1957, it has been cited only 8 times. As a result, upward-bending
curtains have been rediscovered -- albeit in a more general formulation
including shear and viscosity by \cite{Benilov19} (hereinafter, B19). This
paper confirmed the criterion $We<1$ subject to a suitably modified
definition of $We$ for sheared flows. Solutions describing upward-bending
\emph{jets} have been found by \cite{Wallwork01}, but these are strongly
unstable due to the Plateau--Rayleigh instability and, thus, cannot be
observed experimentally.

The results obtained in B19 (and, by association, those of Keller \& Weitz)
have been criticised by \cite{WeinsteinRossRuschakBarlow19}, who put forward
the following claims:

\begin{enumerate}
\item[(1)] For upward-bending curtains, the hyperbolic second-order set
derived in B19 is such that one of the two characteristics corresponds to
waves propagating \emph{upstream} -- i.e., towards the outlet, were the
boundary conditions are set. Thus, \textquotedblleft in accordance with
hyperbolic theory\textquotedblright\ one of the two boundary conditions at
the outlet must be omitted -- namely, the one prescribing the ejection angle
$\alpha _{0}$.

\item[(2)] The omitted boundary condition should be replaced with $\alpha
_{0}=-90^{\circ }$ corresponding to a vertical curtain.
\end{enumerate}

It turns out that (1) is a valid point, whereas (2) is not.

That is, mathematically, there is no reason why a boundary condition cannot
be formulated at a point other than the beginning of a characteristic (as
long as there is only one such point per characteristic). Physically,
however, such a condition would be in conflict with the causality principle,
as it would effectively constrain events occurring in the past.

As for point (2), Weinstein \emph{et al.} justified it by claiming that the
condition $\alpha _{0}=-90^{\circ }$ \textquotedblleft is precisely that
necessary to eliminate the singularity in the curtain\textquotedblright .
However, B19 found regular solutions for \emph{all} values of $\alpha _{0}$,
and earlier in their paper Weinstein \emph{et al.} did not dispute their
existence. One might add that it is inconsistent to use point (1) to discard
one of the boundary conditions and then replace it with a similar one,
leaving their total number exactly the same.

Even more importantly, the replacement of the actual ejection angle with $%
-90^{\circ }$ implies a sharp bend in the curtain near the outlet, and it
cannot be caused by the force of gravity. If it were, the asymptotic models
of both Weinstein \emph{et al.} and B19 would have detected it (as they both
include gravity) -- hence, there would be no need to introduce the turn
through the boundary condition. Nor can it be caused by other hydrodynamic
effects, as they are all isotropic and, thus, cannot make the curtain
`choose' the vertical trajectory regardless of the ejection angle (not to
mention that the only significant ones of these effects -- viscosity and
surface tension -- are also included in the B19 model).

Still, one question remains: if the solutions found in B19 for
upward-bending curtains are indeed physically meaningless, what \emph{really}
happens if liquid is ejected obliquely with $We<1$?

The present paper offers a possible answer to this question through an
analysis of the simplest setting: that of ideal fluid and almost shearless
curtains. The Weber number is assumed to be close to unity, in which case
the phase velocity of the upstream capillary waves is small, so that their
dispersion comes into play. Since the velocity of dispersive waves is not
bounded above, events occurring anywhere in the curtain are immediately
sensed near the outlet, thus resolving the conflict with the causality
principle.

Mathematically, the asymptotic limit $We\approx 1$ invalidates all of the
existing models of oblique liquid curtains %
\citep[including][]{KellerWeitz57,FinnicumWeinsteinRuschak93,Benilov19}: the
leading-order terms almost cancel each other, so the previously-discarded
next-order terms have to be taken into account (as was done by \cite{Ramos03}
for vertical curtains). Once the correct equation for this limit is derived,
one can see that it is not hyperbolic: it involves a first-order time
derivative and a third-order spatial derivative (similarly to the
Korteweg--de Vries equation for gravity-capillary waves in shallow water).
As a result, the whole notion of characteristics becomes irrelevant.

The equation derived will be used to examine steady curtains, and it will be
shown that some of them do bend upwards and rise until all of the liquid's
kinetic energy is used up (and the asymptotic governing equation becomes
inapplicable); other curtains rise until they are truncated by a
self-intersection. There are also solutions describing downward-bending
curtains, both infinite and self-intersecting.

Interestingly, the model seems to `know' that it becomes almost hyperbolic
when the deviation of $We$ from unity is order-one: subcritical ($We<1$)
solutions in this case are highly sensitive to small variations of the
parameters involved, with both upward- and downward-bending curtains often
self-intersecting very near the outlet. One can assume that they are
structurally unstable -- hence, physically meaningless -- and the flow in
this case is unlikely to settle down into a steady curtain. But \emph{%
near-critical} curtains, including subcritical ones, could be observable.
Admittedly, this does not conclusively follow from their existence as steady
solutions of the governing equations (which is shown in this paper), as one
should also prove their stability (which has yet to be done).

The present paper has the following structure: in \S \ref{section 2}, we
formulate the problem and, in \S \ref{section 3}, derive an asymptotic
equation for curtains with a large Froude number and near-unity Weber
number. Steady solutions of this equation are examined in \S \ref{section 4}
and their physical aspects (e.g., how to create an upward-bending curtain in
an experiment), in \S \ref{section 5}.

\section{Formulation of the problem\label{section 2}}

\subsection{Governing equations\label{section 2.1}}

Consider an incompressible ideal fluid of density $\rho $ ejected from an
infinitely long horizontal slot (outlet) of fixed width. Let the flow be
homogeneous in the along-the-outlet direction -- i.e., depend on a single
horizontal coordinate $x$ and the vertical coordinate $z$ (see figure \ref%
{fig1}).

\begin{figure}\vspace*{5mm}\centering
\includegraphics[width=80mm]{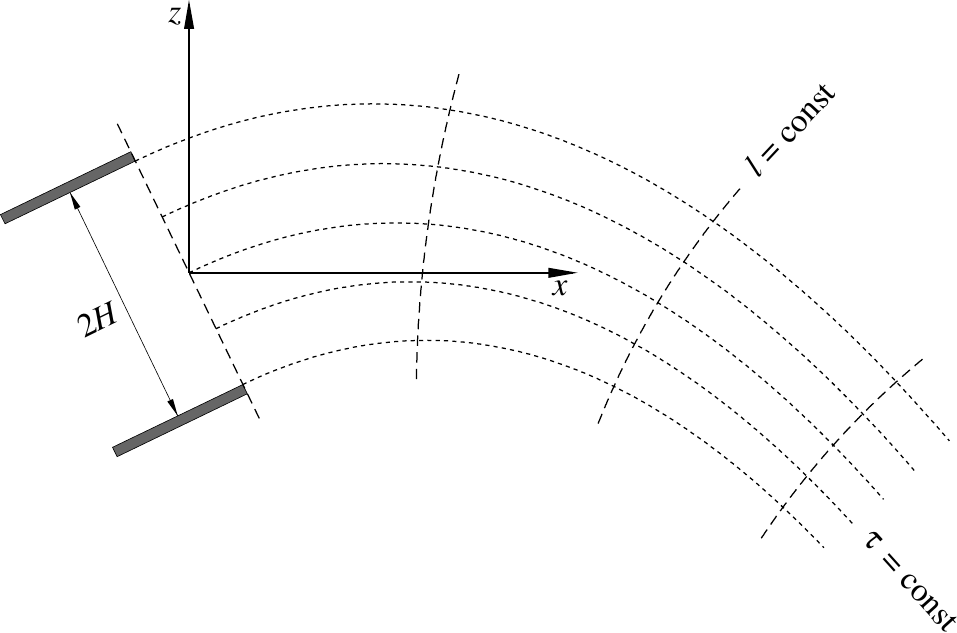}\vspace*{3mm}
\caption{The setting: a two-dimensional liquid curtain ejected from an outlet of width $2H$. $\left( x,z\right) $ are the Cartesian coordinates and $\left( l,\tau \right) $ are the curvilinear coordinates associated with the curtain's centreline (corresponding to $\tau = 0$). \newline\hspace*{6mm} The important quantities not shown in the figure: the local angle $\alpha (l,t)$ between the centreline and the horizontal, its near-outlet value $\alpha _{0} $ (determined by the outlet's orientation -- hence, time-independent), and the near-outlet curvature $\alpha _{0}^{\prime }(t)=\left( \partial \alpha /\partial l\right) _{l=0}$.} \label{fig1}
\end{figure}

Unless the liquid is ejected vertically, the curtain's trajectory is curved
due to gravity, making the Cartesian coordinates awkward to use. It is more
convenient to employ curvilinear coordinates associated with the curtain's
centreline %
\citep{EntovYarin84,Wallwork01,WallworkDecentKingSchulkes02,ShikhmurzaevSisoev17,DecentParauSimmonsUddin18,Benilov19}%
.

Consider curvilinear coordinates $\left( l,\tau \right) $ related to their
Cartesian counterparts by%
\begin{equation*}
x=x(l,\tau ,t),\qquad z=z(l,\tau ,t),
\end{equation*}%
where $t$ is the time. Let the set $\left( l,\tau \right) $ be orthogonal
with a unit Jacobian,%
\begin{equation}
\frac{\partial x}{\partial l}\frac{\partial x}{\partial \tau }+\frac{%
\partial z}{\partial l}\frac{\partial z}{\partial \tau }=0,\qquad \frac{%
\partial x}{\partial l}\frac{\partial z}{\partial \tau }-\frac{\partial x}{%
\partial \tau }\frac{\partial z}{\partial l}=1.  \label{2.1}
\end{equation}%
Since the solution of equations (\ref{2.1}) is not constrained by boundary
conditions, the relationship between $\left( x,z\right) $ and $\left( l,\tau
\right) $ is not unique, leaving one an opportunity to choose the set $%
\left( l,\tau \right) $ that makes the forthcoming calculations simpler.

In what follows, the so-called Lam\'{e} coefficients%
\begin{equation}
h_{l}=\sqrt{\left( \frac{\partial x}{\partial l}\right) ^{2}+\left( \frac{%
\partial z}{\partial l}\right) ^{2}},\qquad h_{\tau }=\sqrt{\left( \frac{%
\partial x}{\partial \tau }\right) ^{2}+\left( \frac{\partial z}{\partial
\tau }\right) ^{2}}  \label{2.2}
\end{equation}%
will be needed. It follows from (\ref{2.1}) that $h_{l}h_{\tau }=1$, i.e.,
the transformation $\left( x,z\right) \rightarrow \left( l,\tau \right) $
preserves the elemental area.

Let the flow be characterised by the velocity components $\left(
u_{l},u_{\tau }\right) $ and the pressure $p$. Representing the
gravitational force by $-\rho g\mathbf{\nabla }z$ ($g$ is the acceleration
due to gravity), one can write the Euler equations in the form%
\begin{multline}
h_{l}\frac{\partial u_{l}}{\partial t}+\left[ u_{l}-\frac{1}{h_{l}}\left(
\frac{\partial x}{\partial l}\frac{\partial x}{\partial t}+\frac{\partial z}{%
\partial l}\frac{\partial z}{\partial t}\right) \right] \left( \frac{%
\partial u_{l}}{\partial l}+\frac{u_{\tau }}{h_{\tau }}\dfrac{\partial h_{l}%
}{\partial \tau }\right) \\
+\frac{h_{l}}{h_{\tau }}\left[ u_{\tau }-\frac{1}{h_{\tau }}\left( \frac{%
\partial x}{\partial \tau }\frac{\partial x}{\partial t}+\frac{\partial z}{%
\partial \tau }\frac{\partial z}{\partial t}\right) \right] \frac{\partial
u_{l}}{\partial \tau } \\
+\frac{u_{\tau }}{h_{\tau }}\left[ \dfrac{\partial x}{\partial l}\frac{%
\partial ^{2}x}{\partial t\,\partial \tau }+\dfrac{\partial z}{\partial l}%
\frac{\partial ^{2}z}{\partial t\,\partial \tau }-\frac{1}{h_{\tau }^{2}}%
\left( \frac{\partial x}{\partial \tau }\frac{\partial x}{\partial t}+\frac{%
\partial z}{\partial \tau }\frac{\partial z}{\partial t}\right) \left(
\dfrac{\partial x}{\partial l}\frac{\partial ^{2}x}{\partial \tau ^{2}}+%
\dfrac{\partial z}{\partial l}\frac{\partial ^{2}z}{\partial \tau ^{2}}%
\right) -u_{\tau }\frac{\partial h_{\tau }}{\partial l}\right] \\
+\dfrac{1}{\rho }\dfrac{\partial p}{\partial l}=-g\dfrac{\partial z}{%
\partial l},  \label{2.3}
\end{multline}%
\begin{multline}
h_{\tau }\frac{\partial u_{\tau }}{\partial t}+\frac{h_{\tau }}{h_{l}}\left[
u_{l}-\frac{1}{h_{l}}\left( \frac{\partial x}{\partial l}\frac{\partial x}{%
\partial t}+\frac{\partial z}{\partial l}\frac{\partial z}{\partial t}%
\right) \right] \frac{\partial u_{\tau }}{\partial l} \\
+\left[ u_{\tau }-\frac{1}{h_{\tau }}\left( \frac{\partial x}{\partial \tau }%
\frac{\partial x}{\partial t}+\frac{\partial z}{\partial \tau }\frac{%
\partial z}{\partial t}\right) \right] \left( \frac{\partial u_{\tau }}{%
\partial \tau }+\frac{u_{l}}{h_{l}}\frac{\partial h_{\tau }}{\partial l}%
\right) \\
+\frac{u_{l}}{h_{l}}\left[ \dfrac{\partial x}{\partial \tau }\frac{\partial
^{2}x}{\partial t\,\partial l}+\dfrac{\partial z}{\partial \tau }\frac{%
\partial ^{2}z}{\partial t\,\partial l}-\frac{1}{h_{l}^{2}}\left( \frac{%
\partial x}{\partial l}\frac{\partial x}{\partial t}+\frac{\partial z}{%
\partial l}\frac{\partial z}{\partial t}\right) \left( \dfrac{\partial x}{%
\partial \tau }\frac{\partial ^{2}x}{\partial l^{2}}+\dfrac{\partial z}{%
\partial \tau }\frac{\partial ^{2}z}{\partial l^{2}}\right) -u_{l}\frac{%
\partial h_{l}}{\partial \tau }\right] \\
+\dfrac{1}{\rho }\dfrac{\partial p}{\partial \tau }=-g\dfrac{\partial z}{%
\partial \tau },  \label{2.4}
\end{multline}%
\begin{equation}
\frac{\partial \left( u_{l}h_{\tau }\right) }{\partial l}+\frac{\partial
\left( u_{\tau }h_{l}\right) }{\partial \tau }=0.  \label{2.5}
\end{equation}%
For a stationary coordinate system (such that $\partial x/\partial
l=\partial x/\partial t=0$), equations (\ref{2.3})--(\ref{2.5}) can be found
in most textbooks \citep[e.g.][]{KochinKibelRoze64}, with the general case
considered in B19.

Let the liquid be bounded by two free surfaces described by equations $\tau
=\tau _{-}(l,t)$ and $\tau =\tau _{+}(l,t)$, where the functions $\tau _{\pm
}(l,\tau )$ satisfy the free-boundary kinematic condition. If written in
terms of time-dependent curvilinear coordinates, it takes the form%
\begin{multline}
\frac{\partial \tau _{\pm }}{\partial t}+\frac{1}{h_{l}}\left[ u_{l}-\dfrac{1%
}{h_{l}}\left( \frac{\partial x}{\partial l}\frac{\partial x}{\partial t}+%
\frac{\partial z}{\partial l}\frac{\partial z}{\partial t}\right) \right]
\dfrac{\partial \tau _{\pm }}{\partial l} \\
-\frac{1}{h_{\tau }}\left[ u_{\tau }-\dfrac{1}{h_{\tau }}\left( \frac{%
\partial x}{\partial \tau }\frac{\partial x}{\partial t}+\frac{\partial z}{%
\partial \tau }\frac{\partial z}{\partial t}\right) \right] =0\qquad \text{if%
}\qquad \tau =\tau _{\pm }.  \label{2.6}
\end{multline}%
The dynamic boundary condition, in turn, assumes that the pressure at a free
boundary is determined by its curvature and surface tension $\sigma $,%
\begin{equation}
p=\mp \sigma \left\{ \frac{\partial }{\partial l}\left[ \frac{\dfrac{h_{\tau
}^{2}}{h_{l}}\dfrac{\partial \tau _{\pm }}{\partial l}}{\sqrt{1+\left(
\dfrac{h_{\tau }}{h_{l}}\dfrac{\partial \tau _{\pm }}{\partial l}\right) ^{2}%
}}\right] -\frac{\partial }{\partial \tau }\left[ \frac{h_{l}}{\sqrt{%
1+\left( \dfrac{h_{\tau }}{h_{l}}\dfrac{\partial \tau _{\pm }}{\partial l}%
\right) ^{2}}}\right] \right\} \qquad \text{if}\qquad \tau =\tau _{\pm }
\label{2.7}
\end{equation}%
(for more details on conditions (\ref{2.6})--(\ref{2.7}), see B19).

Let the outlet's width be $2H$. In this paper, the simplest particular case
is examined, where the streamwise component of the velocity at the outlet is
not sheared (i.e., is independent of $\tau $), while the cross-stream
component is identically zero,%
\begin{equation}
u_{l}=u_{0},\qquad u_{\tau }=0\qquad \text{if}\qquad l=0,\qquad \tau \in
\left( -H,H\right) ,  \label{2.8}
\end{equation}%
where $u_{0}$ may depend on $t$. The outlet conditions for the curtain's
boundaries are, obviously,%
\begin{equation}
\tau _{\pm }=\pm H\qquad \text{if}\qquad l=0.  \label{2.9}
\end{equation}%
Given an initial condition and a specific solution of the coordinate
equations (\ref{2.1}), the boundary-value problem (\ref{2.2})--(\ref{2.9})
(presumably uniquely) determines $p$, $u_{l}$, $u_{\tau }$, and $\tau _{\pm
} $.

It is convenient to identify the curtain's centreline with the curve $\tau
=0 $. To do so, require that%
\begin{equation}
\tau _{\pm }=\pm W,  \label{2.10}
\end{equation}%
where $W(l,t)$ is the curtain's half-width.

\subsection{The scaling\label{section 2.2}}

There are two nondimensional parameters in the problem at hand. The Froude
number%
\begin{equation*}
Fr=\frac{u_{0}^{2}}{gH},
\end{equation*}%
reflects the balance of inertia and gravity, and the Weber number%
\begin{equation*}
We=\frac{\rho Hu_{0}^{2}}{\sigma }
\end{equation*}%
that of inertia and surface tension. This paper is concerned the limit $%
We\approx 1$, which implies that the velocity should be scaled by%
\begin{equation}
U=\left( \frac{\sigma }{\rho H}\right) ^{1/2}.  \label{2.11}
\end{equation}%
In B19, the curtain's spatial scale $L$ was derived from the assumption that
the centrifugal force (due to the curtain's curvature), surface tension, and
gravity are in balance -- which implies that $L=H\,Fr$, so that the
slender-curtain approximation is valid if $Fr\gg 1$.

If, however, $We\rightarrow 1$, the leading-order surface tension and
centrifugal force tend to cancel each other. As a result, gravity remains
unopposed and makes the curtain bend much more steeply (see figure 2 in B19).

To determine $L$ for the case $We\approx 1$, one should balance gravity with
\emph{higher-order corrections} to centrifugal force and surface tension --
which have to be calculated first, of course -- which is, however,
impossible without a conjecture as to what $L$ actually is. This task is
exacerbated even more by the fact that the next-to-leading-order corrections
do contribute to the eventual asymptotic equation, forcing one to delve into
yet another order.

Thus, $L$ was effectively determined through a trial-and-error approach, and
it has turned out that a consistent asymptotic theory can be derived if%
\begin{equation}
L=\frac{H}{\epsilon },  \label{2.12}
\end{equation}%
where $\epsilon $ is the cube root of the Bond number,%
\begin{equation}
\epsilon =\left( \frac{\rho gH^{2}}{\sigma }\right) ^{1/3},  \label{2.13}
\end{equation}%
or, equivalently, $\epsilon =\left( We/Fr\right) ^{1/3}\approx Fr^{-1/3}$.

In the present paper, $\epsilon $ is assumed small -- hence, $L\gg H$. The
larger scale, $L$, will be used to nondimensionalise the Cartesian
coordinates $\left( x,z\right) $ and the streamwise variable $l$, whereas
the cross-stream variable $\tau $ and the curtain's half-width $W$ will be
nondimensionalised by $H$.

The following nondimensional variables will be used:%
\begin{equation}
\left.
\begin{array}{c}
l_{nd}=\dfrac{l}{L},\qquad \tau _{nd}=\dfrac{\tau }{H},\qquad t_{nd}=\dfrac{t%
}{T},\qquad x_{nd}=\dfrac{x}{L},\qquad z_{nd}=\dfrac{z}{L},\medskip  \\
\left( u_{l}\right) _{nd}=\dfrac{u_{l}}{U},\qquad \left( u_{\tau }\right)
_{nd}=\dfrac{u_{\tau }}{\epsilon ^{2}U},\qquad p_{nd}=\dfrac{p}{P},\qquad
W_{nd}=\dfrac{W}{H}.%
\end{array}%
\right\}   \label{2.14}
\end{equation}%
The pressure scale $P$ is such that the centrifugal force is, to leading
order, balanced by the capillary-pressure gradient, which implies%
\begin{equation}
P=\frac{\rho gH}{\epsilon ^{2}}.  \label{2.15}
\end{equation}%
The time scale $T$ is determined by the balance of the Coriolis force and
gravity -- i.e., the time derivatives of $x$ and $z$ in equation (\ref{2.4})
and its right-hand side, respectively -- which amounts to%
\begin{equation}
T=\frac{H}{\epsilon ^{3}U}.  \label{2.16}
\end{equation}%
Rewriting (\ref{2.1})--(\ref{2.7}) in terms of variables (\ref{2.14})--(\ref%
{2.16}), taking into account (\ref{2.11})--(\ref{2.13}), and omitting the
subscript $_{nd}$, one obtains%
\begin{equation}
\frac{\partial x}{\partial l}\frac{\partial x}{\partial \tau }+\frac{%
\partial z}{\partial l}\frac{\partial z}{\partial \tau }=0,\qquad \frac{1}{%
\epsilon }\left( \frac{\partial x}{\partial l}\frac{\partial z}{\partial
\tau }-\frac{\partial x}{\partial \tau }\frac{\partial z}{\partial l}\right)
=1,  \label{2.17}
\end{equation}%
\begin{equation}
h_{l}=\sqrt{\left( \frac{\partial x}{\partial l}\right) ^{2}+\left( \frac{%
\partial z}{\partial l}\right) ^{2}},\qquad h_{\tau }=\frac{1}{\epsilon }%
\sqrt{\left( \frac{\partial x}{\partial \tau }\right) ^{2}+\left( \frac{%
\partial z}{\partial \tau }\right) ^{2}},  \label{2.18}
\end{equation}%
\begin{multline}
\epsilon h_{l}\frac{\partial u_{l}}{\partial t}+\left[ u_{l}-\frac{\epsilon
^{2}}{h_{l}}\left( \frac{\partial x}{\partial l}\frac{\partial x}{\partial t}%
+\frac{\partial z}{\partial l}\frac{\partial z}{\partial t}\right) \right]
\left( \frac{1}{\epsilon }\frac{\partial u_{l}}{\partial l}+\frac{u_{\tau }}{%
h_{\tau }}\dfrac{\partial h_{l}}{\partial \tau }\right)  \\
+\frac{h_{l}}{h_{\tau }}\left[ u_{\tau }-\frac{1}{\epsilon h_{\tau }}\left(
\frac{\partial x}{\partial \tau }\frac{\partial x}{\partial t}+\frac{%
\partial z}{\partial \tau }\frac{\partial z}{\partial t}\right) \right]
\frac{\partial u_{l}}{\partial \tau } \\
+\frac{\epsilon ^{2}u_{\tau }}{h_{\tau }}\left[ \dfrac{\partial x}{\partial l%
}\frac{\partial ^{2}x}{\partial t\,\partial \tau }+\dfrac{\partial z}{%
\partial l}\frac{\partial ^{2}z}{\partial t\,\partial \tau }-\frac{1}{%
\epsilon ^{2}h_{\tau }^{2}}\left( \frac{\partial x}{\partial \tau }\frac{%
\partial x}{\partial t}+\frac{\partial z}{\partial \tau }\frac{\partial z}{%
\partial t}\right) \left( \dfrac{\partial x}{\partial l}\frac{\partial ^{2}x%
}{\partial \tau ^{2}}+\dfrac{\partial z}{\partial l}\frac{\partial ^{2}z}{%
\partial \tau ^{2}}\right) -\epsilon u_{\tau }\frac{\partial h_{\tau }}{%
\partial l}\right]  \\
+\frac{\partial p}{\partial l}=-\epsilon \dfrac{\partial z}{\partial l},
\label{2.19}
\end{multline}%
\begin{multline}
\epsilon ^{4}h_{\tau }\frac{\partial u_{\tau }}{\partial t}+\frac{\epsilon
^{2}h_{\tau }}{h_{l}}\left[ u_{l}-\frac{\epsilon ^{2}}{h_{l}}\left( \frac{%
\partial x}{\partial l}\frac{\partial x}{\partial t}+\frac{\partial z}{%
\partial l}\frac{\partial z}{\partial t}\right) \right] \frac{\partial
u_{\tau }}{\partial l} \\
+\epsilon ^{3}\left[ u_{\tau }-\frac{1}{\epsilon h_{\tau }}\left( \frac{%
\partial x}{\partial \tau }\frac{\partial x}{\partial t}+\frac{\partial z}{%
\partial \tau }\frac{\partial z}{\partial t}\right) \right] \left( \frac{%
\partial u_{\tau }}{\partial \tau }+\frac{u_{l}}{\epsilon h_{l}}\frac{%
\partial h_{\tau }}{\partial l}\right)  \\
+\frac{u_{l}}{h_{l}}\left[ \epsilon \left( \dfrac{\partial x}{\partial \tau }%
\frac{\partial ^{2}x}{\partial t\,\partial l}+\dfrac{\partial z}{\partial
\tau }\frac{\partial ^{2}z}{\partial t\,\partial l}\right) -\frac{\epsilon }{%
h_{l}^{2}}\left( \frac{\partial x}{\partial l}\frac{\partial x}{\partial t}+%
\frac{\partial z}{\partial l}\frac{\partial z}{\partial t}\right) \left(
\dfrac{\partial x}{\partial \tau }\frac{\partial ^{2}x}{\partial l^{2}}+%
\dfrac{\partial z}{\partial \tau }\frac{\partial ^{2}z}{\partial l^{2}}%
\right) -\frac{u_{l}}{\epsilon }\frac{\partial h_{l}}{\partial \tau }\right]
\\
+\frac{\partial p}{\partial \tau }=-\epsilon \dfrac{\partial z}{\partial
\tau },  \label{2.20}
\end{multline}%
\begin{equation}
\frac{1}{\epsilon }\frac{\partial \left( u_{l}h_{\tau }\right) }{\partial l}+%
\frac{\partial \left( u_{\tau }h_{l}\right) }{\partial \tau }=0,
\label{2.21}
\end{equation}%
\begin{multline}
\epsilon \frac{\partial W}{\partial t}+\dfrac{1}{\epsilon h_{l}}\left[ u_{l}-%
\dfrac{\epsilon ^{2}}{h_{l}}\left( \frac{\partial x}{\partial l}\frac{%
\partial x}{\partial t}+\frac{\partial z}{\partial l}\frac{\partial z}{%
\partial t}\right) \right] \dfrac{\partial W}{\partial l} \\
\mp \frac{1}{h_{\tau }}\left[ u_{\tau }-\dfrac{1}{\epsilon h_{\tau }}\left(
\frac{\partial x}{\partial \tau }\frac{\partial x}{\partial t}+\frac{%
\partial z}{\partial \tau }\frac{\partial z}{\partial t}\right) \right]
=0\qquad \text{if}\qquad \tau =\pm W,  \label{2.22}
\end{multline}%
\begin{equation}
p=-\epsilon \frac{\partial }{\partial l}\left[ \frac{\dfrac{h_{\tau }^{2}}{%
h_{l}}\dfrac{\partial W}{\partial l}}{\sqrt{1+\left( \dfrac{\epsilon h_{\tau
}}{h_{l}}\dfrac{\partial W}{\partial l}\right) ^{2}}}\right] \pm \frac{1}{%
\epsilon }\frac{\partial }{\partial \tau }\left[ \frac{h_{l}}{\sqrt{1+\left(
\dfrac{\epsilon h_{\tau }}{h_{l}}\dfrac{\partial W}{\partial l}\right) ^{2}}}%
\right] \qquad \text{if}\qquad \tau =\pm W,  \label{2.23}
\end{equation}%
When nondimensionalising the boundary conditions (\ref{2.8})--(\ref{2.9}),
it is convenient to introduce the nondimensional `excess injection velocity'
$v_{0}$ such that%
\begin{equation}
u_{0}=U\left( 1+\epsilon ^{2}v_{0}\right) .  \label{2.24}
\end{equation}%
Then, (\ref{2.8}) becomes%
\begin{equation}
u_{l}=1+\epsilon ^{2}v_{0},\qquad u_{\tau }=0\qquad \text{if}\qquad
l=0,\qquad \tau \in \left( -1,1\right) .  \label{2.25}
\end{equation}%
Finally, the nondimensional version of the boundary condition (\ref{2.9})--(%
\ref{2.10}), is%
\begin{equation}
W=1\qquad \text{if}\qquad l=0.  \label{2.26}
\end{equation}

\subsection{How should the curvilinear coordinates be chosen?\label{section
2.3}}

As stated before, the curvilinear coordinates $\left( l,\tau \right) $ are
associated with the curtain's centreline -- but this association has not
been reflected in the general equations (\ref{2.17}) relating $\left( l,\tau
\right) $ to the Cartesian coordinates.

To single out the desired solution of (\ref{2.17}), introduce the
centreline's Cartesian coordinates $x=\bar{x}(l,t)$ and $z=\bar{z}(l,t)$,
and the local angle $\alpha (l,t)$ between the centreline and the
horizontal. Let $l$ be the centreline's arclength, which implies%
\begin{equation}
\frac{\partial \bar{x}}{\partial l}=\cos \alpha ,\qquad \frac{\partial \bar{z%
}}{\partial l}=\sin \alpha ,  \label{2.27}
\end{equation}%
\begin{equation}
\bar{x}=0,\qquad \bar{z}=0\qquad \text{if}\qquad l=0.  \label{2.28}
\end{equation}%
For a given $\alpha (l,t)$, (\ref{2.27})--(\ref{2.28}) uniquely determine $%
\bar{x}(l,t)$ and $\bar{z}(l,t)$.

Now, seek a solution (\ref{2.17}) in the form of a series in powers of $%
\left( \epsilon \tau \right) $, with the zero-order terms in the expansions
of $x$ and $z$ being $\bar{x}(l)$ and $\bar{z}(l)$, respectively. After
straightforward algebra (for more detail, see B19), one obtains%
\begin{equation}
x=\bar{x}-\epsilon \tau \sin \alpha -\frac{(\epsilon \tau )^{2}}{2}\frac{%
\partial \alpha }{\partial l}\sin \alpha -(\epsilon \tau )^{3}\left[ \frac{1%
}{6}\frac{\partial ^{2}\alpha }{\partial l^{2}}\cos \alpha +\frac{1}{2}%
\left( \frac{\partial \alpha }{\partial l}\right) ^{2}\sin \alpha \right] +%
\mathcal{O}(\epsilon ^{4}),  \label{2.29}
\end{equation}%
\begin{equation}
z=\bar{z}+\epsilon \tau \cos \alpha +\frac{(\epsilon \tau )^{2}}{2}\frac{%
\partial \alpha }{\partial l}\cos \alpha -(\epsilon \tau )^{3}\left[ \frac{1%
}{6}\frac{\partial ^{2}\alpha }{\partial l^{2}}\sin \alpha -\frac{1}{2}%
\left( \frac{\partial \alpha }{\partial l}\right) ^{2}\cos \alpha \right] +%
\mathcal{O}(\epsilon ^{4}).  \label{2.30}
\end{equation}%
Substitution of these expressions into (\ref{2.18}) yields the following
expressions for the Lam\'{e} coefficients:%
\begin{equation}
h_{l}=1-\epsilon \tau \frac{\partial \alpha }{\partial l}-\frac{(\epsilon
\tau )^{2}}{2}\left( \frac{\partial \alpha }{\partial l}\right)
^{2}-(\epsilon \tau )^{3}\left[ \frac{1}{6}\frac{\partial ^{3}\alpha }{%
\partial l^{3}}+\frac{1}{2}\left( \frac{\partial \alpha }{\partial l}\right)
^{3}\right] +\mathcal{O}(\epsilon ^{4}),  \label{2.31}
\end{equation}%
\begin{equation}
h_{\tau }=1+\epsilon \tau \frac{\partial \alpha }{\partial l}+\frac{%
3(\epsilon \tau )^{2}}{2}\left( \frac{\partial \alpha }{\partial l}\right)
^{2}+(\epsilon \tau )^{3}\left[ \frac{1}{6}\frac{\partial ^{3}\alpha }{%
\partial l^{3}}+\frac{5}{2}\left( \frac{\partial \alpha }{\partial l}\right)
^{3}\right] +\mathcal{O}(\epsilon ^{4}).  \label{2.32}
\end{equation}

\section{Asymptotic analysis\label{section 3}}

The sheer size of the governing equations makes their analysis cumbersome.
To mitigate this problem, the asymptotic results and their physical meaning
will be summarised first, in \S \ref{section 3.1}, with the technicalities
described in \S \S \ref{section 3.2}--\ref{section 3.5}.

\subsection{Summary\label{section 3.1}}

All characteristics of the flow can be related to the curtain's coordinates $%
\bar{x}(l,t)$ and $\bar{z}(l,t)$, and the local angle $\alpha (l,t)$ between
its centreline and the horizontal. The former satisfy (\ref{2.27})--(\ref%
{2.28}), and the latter is governed by the following asymptotic equation:%
\begin{equation}
\frac{\partial \alpha }{\partial t}+\frac{\partial \alpha }{\partial l}\left[
v_{0}-\frac{1}{2}\bar{z}+\frac{1}{12}\left( \frac{\partial \alpha }{\partial
l}\right) ^{2}-\frac{1}{6}\alpha _{0}^{\prime 2}\right] +\frac{1}{6}\frac{%
\partial ^{3}\alpha }{\partial l^{3}}=-\frac{1}{2}\cos \alpha ,  \label{3.1}
\end{equation}%
where $v_{0}$ is the excess injection velocity [see (\ref{2.25})] and%
\begin{equation}
\alpha _{0}^{\prime }=\left( \frac{\partial \alpha }{\partial l}\right)
_{l=0}  \label{3.2}
\end{equation}%
is the curtain's curvature near the outlet. This characteristic plays an
important role in the curtain's global dynamics.

Equation (\ref{3.1}) requires two boundary conditions at the outlet, with
one of these prescribing the ejection angle,%
\begin{equation}
\alpha =\alpha _{0}\qquad \text{if}\qquad l=0,  \label{3.3}
\end{equation}%
where $\alpha _{0}$ may vary with $t$. An additional boundary condition
follows from the analysis of a near-outlet boundary layer: the solution
there matches the global solution only if the latter satisfies%
\begin{equation}
\frac{\partial ^{2}\alpha }{\partial l^{2}}=0\qquad \text{if}\qquad l=0.
\label{3.4}
\end{equation}%
The set comprising (\ref{3.1})--(\ref{3.4}) and (\ref{2.27})--(\ref{2.28})
is the desired asymptotic model for liquid curtains with a large Froude
number and a near-unity Weber number.

The following comments should helpful to understand the physical meaning of
the asymptotic model.

\begin{itemize}
\item Equation (\ref{3.1}) is, essentially, the cross-stream momentum
equation integrated across the curtain, and so each of its terms can be
interpreted physically:

\begin{itemize}
\item The time derivative represents the Coriolis force -- which is a cross
product of the fluid velocity by the angular velocity of the coordinate
frame. The former is, to leading order, unity and $\partial \alpha /\partial
t$ is the latter.

\item The term $v_{0}-\frac{1}{2}\bar{z}$ is an approximate expression for
the excess velocity affected (through the energy conservation) by the local
height. Physically, it is the phase velocity of upstream-propagating
capillary waves.

\item The right-hand side of (\ref{3.1}) represents the force of gravity.

\item The remaining terms represent the centrifugal force and capillary
pressure gradient (since they both depend on the curtain's curvature, it is
impossible to match each of the corresponding terms to a single effect).
\end{itemize}

\item The presence of $v_{0}$ and $\alpha _{0}^{\prime }$ among the
coefficients of equation (\ref{3.1}) makes the curtain dynamics non-local,
as the effect of what happens near the outlet is \emph{immediately} sensed
downstream. The non-locality is a result of the difference between the
(slow) timescale of the curtain evolution and the (fast) timescale of fluid
particles passing through the region where equation (\ref{3.1}) applies.

\item The derived model governs \emph{sinuous} oscillations of curtains,
whereas \emph{varicose} oscillations
\citep[which are also present in the
exact model -- see][]{BenilovBarrosObrien16} have been scaled out. The
latter are still generated in a boundary layer near the outlet (as shown in
Appendix \ref{Appendix A}), but their small amplitude and short wavelength
make their impact on the global dynamics negligible.

\item It can be demonstrated that%
\begin{equation*}
We=\left( 1+\epsilon ^{2}v_{0}\right) ^{2},
\end{equation*}%
i.e., the deviation of $We$ from unity is controlled by the excess injection
velocity $v_{0}$.\\*[0pt]
\hspace*{0.6cm}Curtains with negative (positive) $v_{0}$ will be referred to
as subcritical (supercritical).
\end{itemize}

\subsection{Derivation of equation (\protect\ref{3.1})\label{section 3.2}}

The solution of set (\ref{2.19})--(\ref{2.26}), (\ref{2.29})--(\ref{2.32})
will be sought in the form%
\begin{equation}
\left.
\begin{array}{c}
u_{l}=1+\epsilon u_{l}^{(1)}+\epsilon ^{2}u_{l}^{(2)}...,\qquad u_{\tau
}=u_{l}^{(0)}+\epsilon u_{l}^{(1)}...,\qquad p=p^{(0)}+\epsilon
p^{(1)}...,\medskip \\
W=1+\epsilon W^{(1)}+\epsilon ^{2}W^{(2)}...~.%
\end{array}%
\right\}  \label{3.5}
\end{equation}%
The fact that the leading-order $u_{l}$ and $W$ are constant suggests that
the described dynamics mainly occurs near the outlet, so that the change in
the potential energy is too small to alter the fluid velocity and,
consequently, the curtain's width. The same fact also implies that the
curtain's evolution mainly consists in the centreline changing its shape.

Unlike the physical variables, those associated with the coordinate system
do not have to be expanded. Higher-order corrections for $\bar{x}$, $\bar{z}$%
, and $\alpha $ need to be introduced only if one intends to derive
asymptotic equations for them, and I do not.

To \underline{leading order}, (\ref{2.19})--(\ref{2.26}) and (\ref{2.29})--(%
\ref{2.32}) yield%
\begin{equation}
\frac{\partial u_{l}^{(1)}}{\partial l}+\frac{\partial p^{(0)}}{\partial l}%
=0,  \label{3.6}
\end{equation}%
\begin{equation}
\frac{\partial \alpha }{\partial l}+\frac{\partial p^{(0)}}{\partial \tau }%
=0,  \label{3.7}
\end{equation}%
\begin{equation}
\frac{\partial }{\partial l}\left( u_{l}^{(1)}+\tau \frac{\partial \alpha }{%
\partial l}\right) +\frac{\partial u_{\tau }^{(0)}}{\partial \tau }=0,
\label{3.8}
\end{equation}%
\begin{equation}
\dfrac{\partial W^{(1)}}{\partial l}\mp \left( u_{\tau }^{(0)}+\frac{%
\partial \bar{x}}{\partial t}\sin \alpha -\frac{\partial \bar{z}}{\partial t}%
\cos \alpha \right) =0\qquad \text{if}\qquad \tau =\pm 1,  \label{3.9}
\end{equation}%
\begin{equation}
p^{(0)}=\mp \frac{\partial \alpha }{\partial l}\qquad \text{if}\qquad \tau
=\pm 1.  \label{3.10}
\end{equation}%
\begin{equation}
u_{l}^{(1)}=0,\qquad \text{if}\qquad l=0,\qquad \tau \in \left( -1,1\right) ,
\label{3.11}
\end{equation}%
\begin{equation}
u_{\tau }^{(0)}=0\qquad \text{if}\qquad l=0,\qquad \tau \in \left(
-1,1\right) ,  \label{3.12}
\end{equation}%
\begin{equation}
W^{(1)}=0\qquad \text{if}\qquad l=0.  \label{3.13}
\end{equation}%
Treating $\alpha $ as if it were given, one can deduce from (\ref{3.7}) and (%
\ref{3.10}) that%
\begin{equation}
p^{(0)}=-\tau \frac{\partial \alpha }{\partial l}.  \label{3.14}
\end{equation}%
Substituting $p^{(0)}$ into (\ref{3.6}) and taking into account (\ref{3.11}%
), one obtains\qquad
\begin{equation}
u_{l}^{(1)}=\tau \left( \frac{\partial \alpha }{\partial l}-\alpha
_{0}^{\prime }\right) ,  \label{3.15}
\end{equation}%
where $\alpha _{0}^{\prime }$ is defined by (\ref{3.2}). Next, it follows
from (\ref{3.8})--(\ref{3.9}) and (\ref{3.13}) that%
\begin{equation}
u_{\tau }^{(0)}=\left( 1-\tau ^{2}\right) \frac{\partial ^{2}\alpha }{%
\partial l^{2}}-\frac{\partial \bar{x}}{\partial t}\sin \alpha +\frac{%
\partial \bar{z}}{\partial t}\cos \alpha ,  \label{3.16}
\end{equation}%
\begin{equation}
W^{(1)}=0.  \label{3.17}
\end{equation}%
Treating the \underline{next-to-leading order} in a similar fashion, one
obtains%
\begin{equation}
p^{(1)}=\left( 1-2\tau ^{2}\right) \left( \frac{\partial \alpha }{\partial l}%
\right) ^{2}+\left( \tau ^{2}-1\right) \alpha _{0}^{\prime }\frac{\partial
\alpha }{\partial l},  \label{3.18}
\end{equation}%
\begin{multline}
u_{l}^{(2)}=u_{0}-z+\frac{\partial \bar{x}}{\partial t}\cos \alpha +\frac{%
\partial \bar{z}}{\partial t}\sin \alpha \\
-\left( 1-\frac{3\tau ^{2}}{2}\right) \left( \frac{\partial \alpha }{%
\partial l}\right) ^{2}+\left( 2-\tau ^{2}\right) \alpha _{0}^{\prime }\frac{%
\partial \alpha }{\partial l}-\left( 1+\frac{\tau ^{2}}{2}\right) \alpha
_{0}^{\prime 2},  \label{3.19}
\end{multline}%
\begin{equation}
u_{\tau }^{(1)}=\left( 3\tau -\frac{11}{3}\tau ^{3}\right) \frac{\partial
\alpha }{\partial l}\frac{\partial ^{2}\alpha }{\partial l^{2}}-2\left( \tau
-\frac{\tau ^{3}}{3}\right) \alpha _{0}^{\prime }\frac{\partial ^{2}\alpha }{%
\partial l^{2}}+\tau \sin \alpha ,  \label{3.20}
\end{equation}%
\begin{equation}
W^{(2)}=\bar{z}-\frac{1}{3}\left( \frac{\partial \alpha }{\partial l}\right)
^{2}-\frac{4}{3}\alpha _{0}^{\prime }\frac{\partial \alpha }{\partial l}+%
\frac{5}{3}\alpha _{0}^{\prime 2}.  \label{3.21}
\end{equation}%
In the \underline{next order}, one only needs the cross-stream equation (\ref%
{2.20}) and the capillary-pressure condition (\ref{2.23}), which have the
form%
\begin{multline}
\frac{\partial u_{\tau }^{(0)}}{\partial l}+\frac{\partial \alpha }{\partial
t}-\left( \frac{\partial x}{\partial t}\cos \alpha +\frac{\partial z}{%
\partial t}\sin \alpha \right) \frac{\partial \alpha }{\partial l}+\left(
2u_{l}^{(2)}+u_{l}^{(1)2}\right) \frac{\partial \alpha }{\partial l} \\
+4\tau u_{l}^{(1)}\left( \frac{\partial \alpha }{\partial l}\right) ^{2}+%
\frac{\tau ^{2}}{2}\left[ 8\left( \frac{\partial \alpha }{\partial l}\right)
^{3}+\frac{\partial ^{3}\alpha }{\partial l^{3}}\right] +\frac{\partial
p^{(2)}}{\partial \tau }=-\cos \alpha ,  \label{3.22}
\end{multline}%
\begin{equation}
p^{(2)}\pm \frac{\partial p^{(1)}}{\partial \tau }W^{(1)}=\mp \left[ \frac{3%
}{2}\left( \frac{\partial \alpha }{\partial l}\right) ^{3}+\frac{1}{2}\frac{%
\partial ^{3}\alpha }{\partial l^{3}}\right] \qquad \text{if}\qquad \tau
=\pm 1.  \label{3.23}
\end{equation}%
Observe that (\ref{3.22})--(\ref{3.23}) involve only one unknown, $p^{(2)}$
-- which can be actually eliminated. Integrating (\ref{3.22}) from $\tau =-1$
to $\tau =1$ and taking into account (\ref{3.23}), one obtains%
\begin{multline*}
\int_{-1}^{-1}\frac{\partial u_{\tau }^{(0)}}{\partial l}\mathrm{d}\tau +2%
\frac{\partial \alpha }{\partial t}-2\left( \frac{\partial x}{\partial t}%
\cos \alpha +\frac{\partial z}{\partial t}\sin \alpha \right) \frac{\partial
\alpha }{\partial l} \\
+\frac{\partial \alpha }{\partial l}\int_{-1}^{-1}\left(
2u_{l}^{(2)}+u_{l}^{(1)2}\right) \mathrm{d}\tau +4\left( \frac{\partial
\alpha }{\partial l}\right) ^{2}\int_{-1}^{-1}\tau u_{l}^{(1)}\mathrm{d}\tau
+\frac{1}{2}\left[ 8\left( \frac{\partial \alpha }{\partial l}\right) ^{3}+%
\frac{\partial ^{3}\alpha }{\partial l^{3}}\right] \\
-3\left( \frac{\partial \alpha }{\partial l}\right) ^{3}-\frac{\partial
^{3}\alpha }{\partial l^{3}}-\left[ \left( \frac{\partial p^{(1)}}{\partial
\tau }\right) _{\tau =-1}+\left( \frac{\partial p^{(1)}}{\partial \tau }%
\right) _{\tau =1}\right] W^{(1)}=-2\cos \alpha ,
\end{multline*}%
Substituting in this equality expressions (\ref{3.14})--(\ref{3.23}) for the
lower-order unknowns and evaluating the integrals involved, one obtains
equation (\ref{3.1}) as required.

\subsection{Derivation of condition (\protect\ref{3.4})\label{section 3.3}}

An attentive reader may have noticed that the boundary condition (\ref{3.12}%
) has not been used in the above calculation of the leading-order
cross-stream velocity $u_{\tau }^{(0)}$. As a result, $u_{\tau }^{(0)}$ does
not assume the prescribed value at the outlet: as follows from (\ref{3.16})
and (\ref{2.27}),%
\begin{equation}
u_{\tau }^{(0)}\rightarrow \left( 1-\tau ^{2}\right) \alpha _{0}^{\prime
\prime }\qquad \text{as}\qquad l\rightarrow 0,  \label{3.24}
\end{equation}%
where%
\begin{equation*}
\alpha _{0}^{\prime \prime }=\left( \frac{\partial ^{2}\alpha }{\partial
l^{2}}\right) _{l=0}.
\end{equation*}%
Thus, condition (\ref{3.12}) holds only if $\partial ^{2}\alpha /\partial
l^{2}$ happens to vanish at $l=0$. This discrepancy arises due to the fact
that none of the leading-order equations (\ref{3.6})--(\ref{3.13}) includes $%
\partial u_{\tau }^{(0)}/\partial l$ -- hence, the solution cannot satisfy a
requirement imposed at a fixed $l$.

This discrepancy suggests that a boundary layer exists, with a solution
satisfying the boundary conditions at the outlet and matching the outer
solution [described by expansions (\ref{3.5})] far from the outlet.

To derive the equations describing the boundary layer, one needs to rescale
the streamwise coordinate: instead of $l_{nd}=l/L$, introduce%
\begin{equation}
\left( l_{b}\right) _{nd}=\frac{l}{H}.  \label{3.25}
\end{equation}%
It can be safely assumed that, near the outlet, the curtain does not
experience sharp turns -- hence, in the boundary layer, $\alpha $ can be
treated as a given function determined by the outer region. Expanding $%
\alpha $ in powers of the long-scale coordinate $l$,%
\begin{equation*}
\alpha =\alpha _{0}+l\alpha _{0}^{\prime }+\frac{l^{2}}{2}\alpha
_{0}^{\prime \prime }+\frac{l^{3}}{6}\alpha _{0}^{\prime \prime \prime }+%
\mathcal{O}(l^{4})\qquad \text{as}\qquad l\rightarrow 0
\end{equation*}%
and rewriting the series in terms of $\left( l_{b}\right) _{nd}$, one
obtains (the subscript $_{nd}$ omitted)%
\begin{equation}
\alpha =\alpha _{0}+\epsilon l_{b}\alpha _{0}^{\prime }+\frac{\epsilon
^{2}l_{b}^{2}}{2}\alpha _{0}^{\prime \prime }+\frac{\epsilon ^{3}l_{b}^{3}}{6%
}\alpha _{0}^{\prime \prime \prime }+\mathcal{O}(\epsilon ^{4}).
\label{3.26}
\end{equation}%
Expressions (\ref{2.31})--(\ref{2.32}) for the Lam\'{e} coefficients should
be treated in a similar manner,%
\begin{multline}
h_{l}=1-\tau \left( \epsilon \alpha _{0}^{\prime }+\epsilon ^{2}l_{b}\alpha
_{0}^{\prime \prime }+\frac{\epsilon ^{3}l_{b}^{2}}{2}\alpha _{0}^{\prime
\prime \prime }\right) \\
-\frac{\tau ^{2}}{2}\left( \epsilon ^{2}\alpha _{0}^{\prime 2}+2\epsilon
^{3}l_{b}\alpha _{0}^{\prime }\alpha _{0}^{\prime \prime }\right) -\epsilon
^{3}\tau ^{3}\left( \frac{1}{6}\alpha _{0}^{\prime \prime \prime }+\frac{1}{2%
}\alpha _{0}^{\prime 3}\right) +\mathcal{O}(\epsilon ^{4}),  \label{3.27}
\end{multline}%
\begin{multline}
h_{\tau }=1+\tau \left( \epsilon \alpha _{0}^{\prime }+\epsilon
^{2}l_{b}\alpha _{0}^{\prime \prime }+\frac{\epsilon ^{3}l_{b}^{2}}{2}\alpha
_{0}^{\prime \prime \prime }\right) \\
+\frac{3\tau ^{2}}{2}\left( \epsilon ^{2}\alpha _{0}^{\prime 2}+2\epsilon
^{3}l_{b}\alpha _{0}^{\prime }\alpha _{0}^{\prime \prime }\right) +\epsilon
^{3}\tau ^{3}\left( \frac{1}{6}\alpha _{0}^{\prime \prime \prime }+\frac{5}{2%
}\alpha _{0}^{\prime 3}\right) +\mathcal{O}(\epsilon ^{4}).  \label{3.28}
\end{multline}%
With $\alpha $ being a given function, the curvilinear coordinates no longer
evolve with the flow and, thus, the curve $\tau =0$ does not necessarily
coincide with the centreline. As a result, $\tau _{+}$ and $\tau _{-}$
should be treated as independent functions, not inter-related by constraint (%
\ref{2.10}).

Replacing in nondimensionalisation (\ref{2.14}) the outer coordinate $l_{nd}$
with its inner counterpart (\ref{3.25}) and keeping in mind expansions (\ref%
{3.26})--(\ref{3.28}), one can rewrite the original equations (\ref{2.19})--(%
\ref{2.25}) in the form (the subscript $_{nd}$ omitted)%
\begin{equation}
\frac{u_{l}}{\epsilon ^{2}}\frac{\partial u_{l}}{\partial l_{b}}-\epsilon
\alpha _{0}^{\prime }u_{\tau }+\frac{1}{\epsilon }\frac{\partial p}{\partial
l_{b}}=-\epsilon \sin \alpha _{0}+\mathcal{O}(\epsilon ^{2}),  \label{3.29}
\end{equation}%
\begin{multline}
\epsilon \left( 1+2\epsilon \tau \alpha _{0}^{\prime }\right) \frac{\partial
u_{\tau }}{\partial l_{b}}+u_{l}^{2}\left\{ \alpha _{0}^{\prime }+\epsilon
\left( 2\tau \alpha _{0}^{\prime 2}+l_{b}\alpha _{0}^{\prime \prime }\right) %
\vphantom{\frac{\epsilon ^{2}}{2}}\right. \\
+\left. \frac{\epsilon ^{2}}{2}\left[ \tau ^{2}\left( 8\alpha _{0}^{\prime
3}+\tau ^{2}\alpha _{0}^{\prime \prime \prime }\right) +8\tau l_{b}\alpha
_{0}^{\prime }\alpha _{0}^{\prime \prime }+l_{b}^{2}\alpha _{0}^{\prime
\prime \prime }\right] \right\} +\frac{\partial p}{\partial \tau }=-\epsilon
^{2}\cos \alpha _{0}+\mathcal{O}(\epsilon ^{3}),  \label{3.30}
\end{multline}%
\begin{multline}
\frac{1}{\epsilon ^{2}}\frac{\partial }{\partial l_{b}}\left\{ u_{l}\left[
1+\epsilon \tau \alpha _{0}^{\prime }+\epsilon ^{2}\left( \frac{3\tau ^{2}}{2%
}\alpha _{0}^{\prime 2}+\tau l_{b}\alpha _{0}^{\prime \prime }\right) %
\vphantom{\frac{l_{b}^{2}}{2}}\right. \right. \\
+\left. \left. \epsilon ^{3}\left( \frac{\tau ^{3}}{6}\alpha _{0}^{\prime
\prime \prime }+\frac{5\tau ^{3}}{2}\alpha _{0}^{\prime 3}+3\tau
^{2}l_{b}\alpha _{0}^{\prime }\alpha _{0}^{\prime \prime }+\frac{\tau
l_{b}^{2}}{2}\alpha _{0}^{\prime \prime \prime }\right) \right] \right\} \\
+\frac{\partial }{\partial \tau }\left[ u_{\tau }\left( 1-\epsilon \tau
\alpha _{0}^{\prime }\right) \right] =\mathcal{O}(\epsilon ^{2}),
\label{3.31}
\end{multline}%
\begin{equation}
\frac{1}{\epsilon ^{2}}\dfrac{\partial \tau _{\pm }}{\partial l_{b}}-\left(
1\mp 2\epsilon \alpha _{0}^{\prime }\right) u_{\tau }=\mathcal{O}(\epsilon
^{2})\qquad \text{if}\qquad \tau =\pm 1,  \label{3.32}
\end{equation}%
\begin{multline}
p=\frac{\mp 1-3\epsilon \alpha _{0}^{\prime }}{\epsilon }\dfrac{\partial
^{2}\tau _{\pm }}{\partial l_{b}^{2}}\mp \alpha _{0}^{\prime }-\epsilon
\left( \alpha _{0}^{\prime 2}\pm l_{b}\alpha _{0}^{\prime \prime }\right) \\
\mp \frac{\epsilon ^{2}}{2}\left( 3\alpha _{0}^{\prime 3}+\alpha
_{0}^{\prime \prime \prime }\pm 4l_{b}\alpha _{0}^{\prime }\alpha
_{0}^{\prime \prime }+l_{b}^{2}\alpha _{0}^{\prime \prime \prime }\right) +%
\mathcal{O}(\epsilon ^{3})\qquad \text{if}\qquad \tau =\pm 1.  \label{3.33}
\end{multline}%
These equations were obtained under the following conjectures:%
\begin{equation*}
u_{l}=1+\mathcal{O}(\epsilon ^{2}),\qquad \tau _{\pm }=\pm 1+\mathcal{O}%
(\epsilon ^{2}),
\end{equation*}%
-- accordingly, the solution of the boundary-value problem (\ref{3.29})--(%
\ref{3.33}) should be sought in the form%
\begin{equation}
\left.
\begin{array}{c}
u_{l}=1+\epsilon ^{2}u_{l}^{(2)}...,\qquad u_{\tau }=u_{\tau
}^{(0)}+\epsilon u_{\tau }^{(1)}...,\qquad p=-\alpha _{0}^{\prime }\tau
+\epsilon p^{(1)}...,\medskip \\
\tau _{\pm }=\pm 1+\epsilon ^{2}\tau _{\pm }^{(2)}...~.%
\end{array}%
\right\}  \label{3.34}
\end{equation}%
To leading order, one obtains%
\begin{equation}
\left.
\begin{array}{c}
\dfrac{\partial u_{l}^{(2)}}{\partial l_{b}}+\dfrac{\partial p^{(1)}}{%
\partial l_{b}}=0, \\
\dfrac{\partial u_{\tau }^{(0)}}{\partial l_{b}}+2\tau \alpha _{0}^{\prime
2}+l_{b}\alpha _{0}^{\prime \prime }+\dfrac{\partial p^{(1)}}{\partial \tau }%
=0, \\
\dfrac{\partial u_{l}^{(2)}}{\partial l_{b}}+\tau \alpha _{0}^{\prime \prime
}+\dfrac{\partial u_{\tau }^{(0)}}{\partial \tau }=0,%
\end{array}%
\right\}  \label{3.35}
\end{equation}%
\begin{equation}
\dfrac{\partial \tau _{\pm }^{(2)}}{\partial l_{b}}-u_{\tau }^{(0)}=0,\qquad
p^{(1)}=\mp \dfrac{\partial ^{2}\tau _{\pm }^{(2)}}{\partial l_{b}^{2}}%
-\alpha _{0}^{\prime 2}\mp l_{b}\alpha _{0}^{\prime \prime }\qquad \text{if}%
\qquad \tau =\pm 1,  \label{3.36}
\end{equation}%
\begin{equation}
u_{l}^{(2)}=u_{0},\qquad u_{\tau }^{(0)}=0,\qquad \tau _{\pm }^{(2)}=0\qquad
\text{if}\qquad l_{b}=0.  \label{3.37}
\end{equation}%
(\ref{3.35})--(\ref{3.37}) can be reduced to a boundary-value problem for $%
u_{\tau }^{(0)}$, which can be solved using the Fourier transformation.
Omitting the technicalities (which are similar to those examined in Appendix %
\ref{Appendix A}), one obtains%
\begin{equation}
u_{\tau }^{(0)}=-\alpha _{0}^{\prime \prime }l_{b}^{2}+A^{(0)}\sinh k_{\ast
}\tau \,\sin k_{\ast }l_{b},  \label{3.38}
\end{equation}%
where $A^{(0)}$ is an undetermined constant and $k_{\ast }$ satisfies%
\begin{equation}
\cosh k_{\ast }-k_{\ast }\sinh k_{\ast }=0.  \label{3.39}
\end{equation}%
If solved numerically, equation (\ref{3.39}) yields $k_{\ast }\approx 1.1997$%
.

The term involving $A^{(0)}$ in solution (\ref{3.38}) describes a
short-scale varicose capillary wave coming from infinity -- bouncing off the
outlet -- going back to infinity. Since the \textquotedblleft
infinity\textquotedblright\ here means the \textquotedblleft outer
region\textquotedblright\ and since short waves have been scaled out from
the outer solution, one sets%
\begin{equation}
A^{(0)}=0.  \label{3.40}
\end{equation}%
Comparing the inner solution (\ref{3.38})--(\ref{3.40}) with the inner limit
of the outer solution [given by (\ref{3.24})], one can see that they match
only if $\alpha _{0}^{\prime \prime }=0$. This requirement amounts to the
boundary condition (\ref{3.4}) for the outer solution, as required.

Note that condition (\ref{3.4}) could be derived by forcing the outer $%
u_{\tau }^{(0)}$ to satisfy the boundary condition at the outlet, i.e.,
without considering the boundary layer. A similar problem, however, arises
in the next order: letting $\alpha _{0}^{\prime \prime }=0$ in (\ref{3.20}),
one obtains%
\begin{equation}
u_{\tau }^{(1)}\rightarrow \tau \sin \alpha _{0}\qquad \text{as}\qquad
l\rightarrow 0,  \label{3.41}
\end{equation}%
i.e., $u_{\tau }^{(1)}$ vanishes at the outlet only if the curtain is
ejected horizontally. This discrepancy can only be resolved by examining the
next order of the boundary-layer problem (\ref{3.29})--(\ref{3.33}), as done
in Appendix \ref{Appendix A}.

It is also shown in Appendix \ref{Appendix A} that the boundary-layer
solution describes \emph{varicose} capillary waves.

\subsection{Discussion: conservation laws\label{section 3.5}}

It is instructive to compare equation (\ref{3.1}) to similar asymptotic
models, such as the lubrication theory or shallow water.

Those typically include an equation reflecting the \emph{mass} conservation
law; in addition, if viscosity is neglected, one can rearrange the model's
equations in the form of an \emph{energy} conservation law. Since, for
steady flows, the fluxes of conserved quantities are first integrals, they
are of help when searching for steady-state solutions.

In the present case, however, it is not clear how equation (\ref{3.1}) can
be rewritten as either mass or energy conservation law.

To understand why, assume for simplicity that the curtain is steady -- i.e.,%
\begin{equation}
\frac{\partial \alpha }{\partial t}=\frac{\partial \bar{x}}{\partial t}=%
\frac{\partial \bar{z}}{\partial t}=0.  \label{3.42}
\end{equation}%
It can be shown that, in this case, the exact governing equations (\ref{2.17}%
)--(\ref{2.23}) preserve the nondimensional mass flux,%
\begin{equation*}
F_{m}=\int_{-W}^{W}u_{l}h_{\tau }\,\mathrm{d}\tau ,
\end{equation*}%
and the nondimensional energy flux,%
\begin{equation*}
F_{e}=\int_{-W}^{W}\left[ \frac{1}{2}\left( u_{l}^{2}+\epsilon ^{4}u_{\tau
}\right) +\epsilon p+\epsilon ^{2}z\right] u_{l}h_{\tau }\,\mathrm{d}\tau .
\end{equation*}%
Note that $F_{e}$ does not include a capillary contribution\footnote{%
To understand why, recall that surface energy is proportional to the area of
the free boundary. This implies that the capillary contribution to $F_{e}$
does not vary along a \emph{steady} curtain: if it did, the surface area
between two cross-sections with different fluxes would be varying in time.
One can further show that, for evolving flows, the capillary contribution to
$F_{e}$ is proportional to $\partial W/\partial t$ -- hence, vanishes for
steady flows.}.

Using the outer expansions (\ref{3.5}), (\ref{3.14})--(\ref{3.23}) to
calculate $F_{m}$ and $F_{e}$, and taking into account (\ref{3.42}), one
obtains%
\begin{equation*}
F_{m}=2+\epsilon ^{2}\left( 2u_{0}+\alpha _{0}^{\prime 2}\right) +\mathcal{O}%
(\epsilon ^{3}),\qquad F_{e}=1+\epsilon ^{2}\left( 3u_{0}-\frac{5\alpha
_{0}^{\prime 2}}{6}\right) +\mathcal{O}(\epsilon ^{3}).
\end{equation*}%
Evidently, the mass and energy fluxes do not depend on $\alpha $ and, thus,
are spatially uniform.

In principle, equation (\ref{3.1}) can reflect a higher-order conservation
law, but this seems unlikely: if it did, its steady-state version would have
a first integral -- but I was unable to find it within a reasonable
timeframe.

\section{Steady curtains\label{section 4}}

Letting $\partial \alpha /\partial t=0$ and omitting the overbars above $%
\bar{x}$ and $\bar{z}$, one can reduce the boundary-value problem comprising
(\ref{3.1}), (\ref{2.27}), (\ref{3.2})--(\ref{3.4}), and (\ref{2.28}) to%
\begin{equation}
\frac{\mathrm{d}\alpha }{\mathrm{d}l}\left[ v_{0}-\frac{1}{2}z+\frac{1}{12}%
\left( \frac{\mathrm{d}\alpha }{\mathrm{d}l}\right) ^{2}-\frac{1}{6}\alpha
_{0}^{\prime 2}\right] +\frac{1}{6}\frac{\mathrm{d}^{3}\alpha }{\mathrm{d}%
l^{3}}=-\frac{1}{2}\cos \alpha ,  \label{4.1}
\end{equation}%
\begin{equation}
\frac{\mathrm{d}x}{\mathrm{d}l}=\cos \alpha ,\qquad \frac{\mathrm{d}z}{%
\mathrm{d}l}=\sin \alpha ,  \label{4.2}
\end{equation}%
\begin{equation}
\alpha =\alpha _{0}\qquad \frac{\mathrm{d}\alpha }{\mathrm{d}l}=\alpha
_{0}^{\prime },\qquad \frac{\mathrm{d}^{2}\alpha }{\mathrm{d}l^{2}}=0\qquad
\text{if}\qquad l=0,  \label{4.3}
\end{equation}%
\begin{equation}
x=0,\qquad z=0\qquad \text{if}\qquad l=0.  \label{4.4}
\end{equation}%
Observe that the equation and boundary condition for $x(l)$ decouple from
the rest of the problem and can be solved separately, after $\alpha (l)$ and
$z(l)$ have been found.

\subsection{The difference between upward- and downward-bending curtains
\label{section 4.1}}

Three parameters appear in the boundary-value problem (\ref{4.1})--(\ref{4.4}%
): the ejection angle $\alpha _{0}$, the excess ejection velocity $v_{0}$,
and the curtain's near-outlet curvature $\alpha _{0}^{\prime }$. The first
two are controlled in an experiment -- hence, should be treated as given.
The third parameter, in turn, can\emph{not} be set by the experimentalist --
hence, the mathematician should either treat it as arbitrary (and find a
solution for each value of $\alpha _{0}^{\prime }$ for which a solution
exists) or determine it as part of the solution (the same way eigenvalues
are determined together with the eigenfunctions).

It turns out that the former is the case for the usual, downward-bending
(DB) curtains -- and the latter, for upward-bending (UB) curtains.

To describe a DB curtain, require%
\begin{equation}
\alpha \rightarrow -\frac{\pi }{2}\qquad \text{as}\qquad l\rightarrow \infty
.  \label{4.5}
\end{equation}%
To examine how the solution of (\ref{4.1})--(\ref{4.4}) approaches this
limit, let%
\begin{equation*}
\alpha =-\frac{\pi }{2}+\tilde{\alpha},\qquad z=-l+\Delta _{\infty }+\tilde{z%
},
\end{equation*}%
where%
\begin{equation*}
\Delta _{\infty }=\lim\limits_{l\rightarrow \infty }\left( z+l\right) ,
\end{equation*}%
so that $\tilde{z}\rightarrow 0$ as $l\rightarrow \infty $. Linearising
equation (\ref{4.1}), one obtains%
\begin{equation}
\frac{\mathrm{d}\tilde{\alpha}}{\mathrm{d}l}\left[ v_{0}-\frac{1}{2}\left(
-l+\Delta _{\infty }\right) -\frac{1}{6}\alpha _{0}^{\prime 2}\right] +\frac{%
1}{6}\frac{\mathrm{d}^{3}\tilde{\alpha}}{\mathrm{d}l^{3}}=-\frac{1}{2}\tilde{%
\alpha},  \label{4.6}
\end{equation}%
whereas the equation for $\tilde{z}$ decouples from the above and is
unimportant.

Let the general solution of (\ref{4.6}) be%
\begin{equation*}
\tilde{\alpha}=C_{1}\,\tilde{\alpha}_{1}(l)+C_{2}\,\tilde{\alpha}%
_{2}(l)+C_{3}\,\tilde{\alpha}_{3}(l),
\end{equation*}%
where $C_{1,2,3}$ are arbitrary constants. The linearly independent
solutions $\tilde{\alpha}_{1,2,3}(l)$ can be fixed by their large-$l$
asymptotics,%
\begin{equation*}
\tilde{\alpha}_{1}\sim \tilde{l}^{-1},\qquad \tilde{\alpha}_{2}\sim \tilde{l}%
^{-1/4}\sin \frac{2\tilde{l}^{3/2}}{3^{1/2}},\qquad \tilde{\alpha}_{3}\sim
\tilde{l}^{-1/4}\cos \frac{2\tilde{l}^{3/2}}{3^{1/2}}\qquad \text{as}\qquad
l\rightarrow \infty ,
\end{equation*}%
where%
\begin{equation*}
\tilde{l}=l-\Delta _{\infty }+2u_{0}-\frac{1}{3}\alpha _{0}^{\prime 2}.
\end{equation*}%
Most importantly, $\tilde{\alpha}_{1,2,3}$ all decay as $l\rightarrow \infty
$ -- which means that the point $\alpha =-\pi /2$ is an attractor. As a
result, solutions are likely to exist for a \emph{range} of $\alpha
_{0}^{\prime }$ (not just a set of discrete values): one can `shoot' a
solution of (\ref{4.1})--(\ref{4.4}) from the outlet with a value of $\alpha
_{0}^{\prime }$ from the allowed range, and this solution will end up at the
attractor. Thus, there is no need to seek new solutions by parametric
continuation from the already-found ones.

UB curtains, in turn, imply that%
\begin{equation}
\alpha \rightarrow \frac{\pi }{2}\qquad \text{as}\qquad l\rightarrow \infty ,
\label{4.7}
\end{equation}%
and the large-$l$ analysis yields%
\begin{equation}
z\sim l+\Delta _{\infty },\qquad \alpha \sim \frac{\pi }{2}+C_{1}\,\tilde{%
\alpha}_{1}+C_{2}\,\tilde{\alpha}_{2}+C_{3}\,\tilde{\alpha}_{3}\qquad \text{%
as}\qquad l\rightarrow \infty ,  \label{4.8}
\end{equation}%
where%
\begin{equation}
\tilde{\alpha}_{1}\sim \tilde{l}^{-1},\qquad \tilde{\alpha}_{2}\sim \tilde{l}%
^{-1/4}\exp \frac{2\tilde{l}^{3/2}}{3^{1/2}},\qquad \tilde{\alpha}_{3}\sim
\tilde{l}^{-1/4}\exp \left( -\frac{2\tilde{l}^{3/2}}{3^{1/2}}\right) \qquad
\text{as}\qquad l\rightarrow \infty ,  \label{4.9}
\end{equation}%
\begin{equation}
\tilde{l}=l+z_{\infty }-2u_{0}+\frac{1}{3}\alpha _{0}^{\prime 2}.
\label{4.10}
\end{equation}%
Evidently, $\tilde{\alpha}_{2}$ grows as $l\rightarrow \infty $ -- hence,
the point $\alpha =\pi /2$ is \emph{not} an attractor. As a result, problem (%
\ref{4.1})--(\ref{4.4}), (\ref{4.7}) may have a solution only for discrete
values of $\alpha _{0}^{\prime }$, such that $C_{2}$ in asymptotic (\ref{4.8}%
) vanishes.

\subsection{Numerical results\label{section 4.2}}

The boundary-value problem (\ref{4.1})--(\ref{4.4}) was solved numerically.
The following results have been obtained.\smallskip

\begin{figure}\vspace*{5mm}
\includegraphics[width=\textwidth]{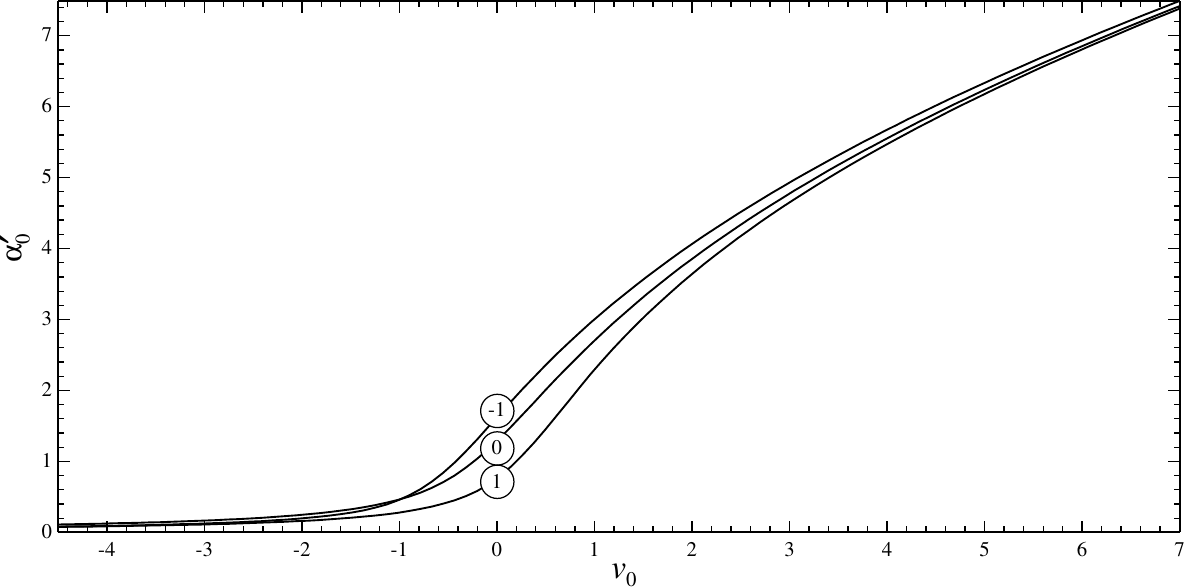}\vspace*{3mm}
\caption{The near-outlet curvature $\alpha _{0}^{\prime }$ of upward-bending curtains \emph{vs} the excess ejection velocity $v_{0}$. The curves marked by ($\pm 1$) and ($0$) correspond to $\alpha _{0}=\pm \pi /4$ and $\alpha _{0}=0$, respectively.}\label{fig2}
\end{figure}

\begin{figure}\vspace*{5mm}
\includegraphics[width=\textwidth]{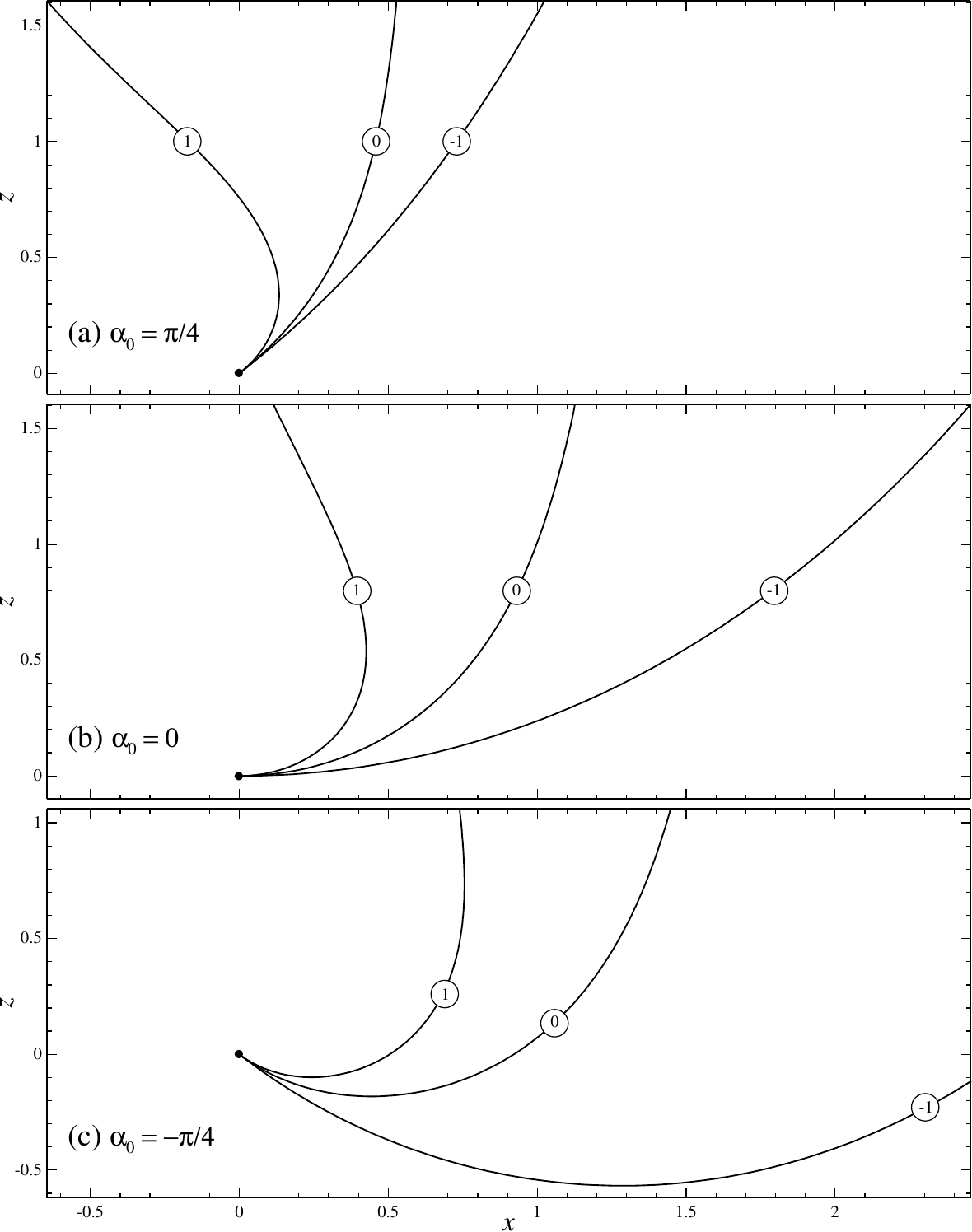}\vspace*{3mm}
\caption{Examples of upward-bending curtains, as described by the boundary-value problem (\ref{4.1})--(\ref{4.4}), (\ref{4.7}). The values of the ejection angle $\alpha _{0}$ are indicated in the corresponding panels,
the values of $v_{0}$ mark the corresponding curves.}\label{fig3}
\end{figure}

(1) The numerical method for the UB problem [comprising (\ref{4.1})--(\ref%
{4.4}) and (\ref{4.7})] is described in Appendix \ref{Appendix B}. The
parameter plane $\left( \alpha _{0},v_{0}\right) $ has been thoroughly
trawled, and it has turned out that, for a given pair $\left( \alpha
_{0},v_{0}\right) $, a UB curtain exists only for a \emph{single} value of $%
\alpha _{0}^{\prime }$. The dependence of $\alpha _{0}^{\prime }$ on $v$ and
$\alpha _{0}$ is illustrated in figure \ref{fig2}, and examples of UB
curtains are shown in figure \ref{fig3}.

\begin{figure}\vspace*{5mm}
\includegraphics[width=\textwidth]{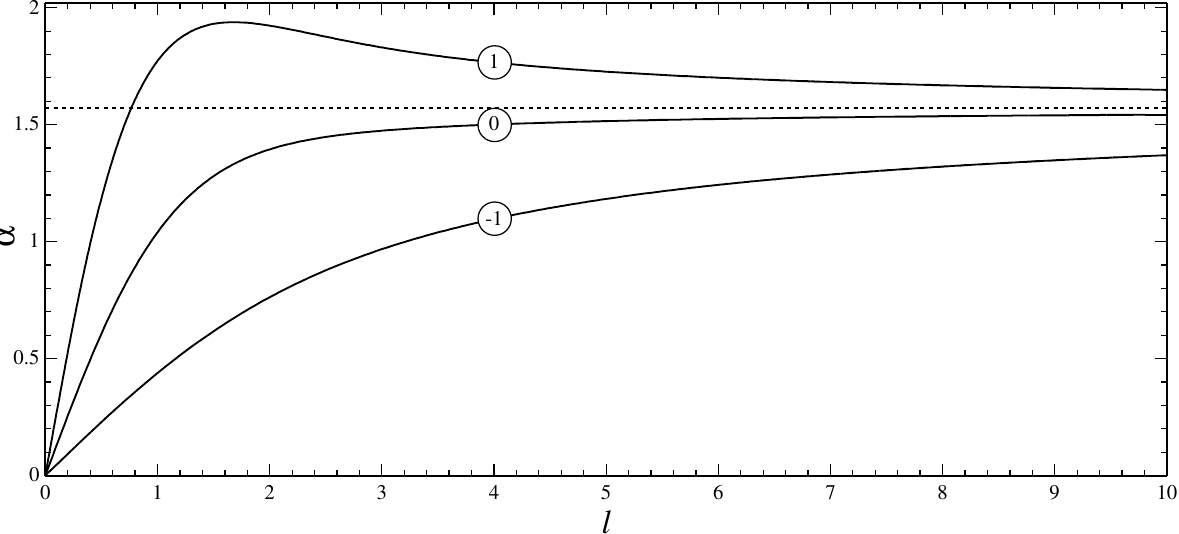}\vspace*{3mm}
\caption{The local angle $\alpha $ between the curtain and the horizontal \emph{vs} the centreline's arclength $l$, for the examples shown in figure \ref{fig3}(b) ($\alpha _{0}=0$, the values of $v_{0}$ mark the corresponding curves). Observe that curve (1) overshoots the vertical direction (shown by the dotted line) and comes back to it after an inflection point.}\label{fig4}
\end{figure}

The mere fact of existence of gravity-defying flows is highly
counter-intuitive, but this is not the only paradoxical feature of the
solutions found. One would expect faster curtains (those with larger $v_{0}$%
) to be straighter than their small-$v_{0}$ counterparts -- but, in reality,
it is the other way around. This tendency is visible in the examples in
figure \ref{fig3}, and is quantified in figure \ref{fig2} (observe the
growth of the near-outlet curvature $\alpha _{0}^{\prime }$ with increasing $%
v_{0}$). Furthermore, sufficiently fast curtains are so curved that they
overshoot the vertical direction and come back to it after an inflection
point (this pattern is not evident from figure \ref{fig3}, but it is clearly
visible in figure \ref{fig4}). It should be emphasized, however, that
solutions with large $\alpha _{0}^{\prime }$ may violate the slender-curtain
approximation. Recalling how the problem was nondimensionalised, one can
show that the applicability condition of the solutions found is $\left\vert
\alpha _{0}^{\prime }\right\vert \ll \epsilon ^{-1}$.

The dependence of UB curtains on the ejection angle $\alpha _{0}$ does not
seem to be essential, as the trajectories of curtains with different values
of $\alpha _{0}$ are qualitatively similar. Observe also that the curves
with different $\alpha _{0}$ in figure \ref{fig2} are close to each other.
One can expect a different behaviour only in the limit $\alpha
_{0}\rightarrow -\pi /2$ (nearly vertical curtains), which is not considered
here.\smallskip

\begin{figure}\vspace*{5mm}
\includegraphics[width=\textwidth]{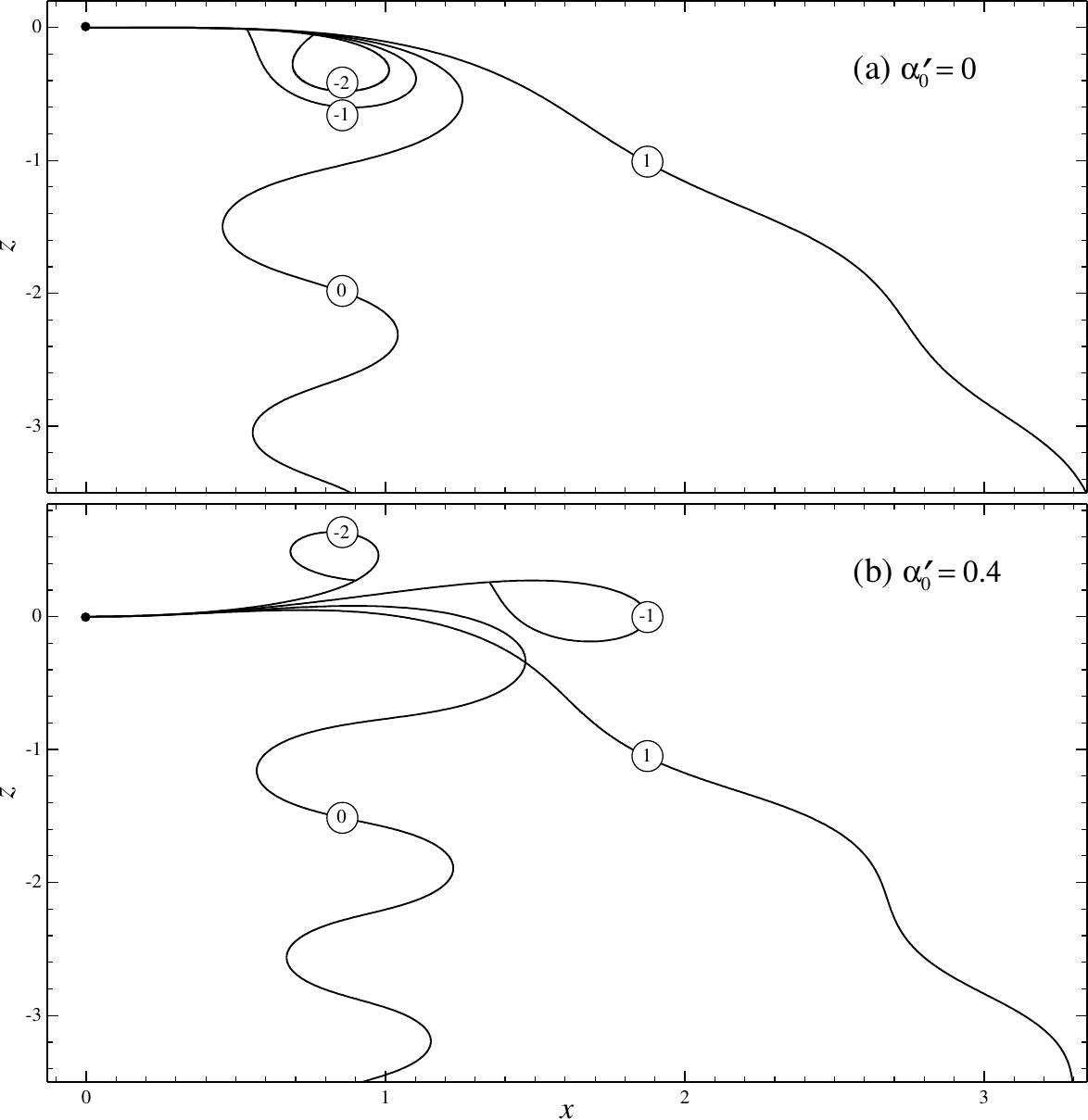}\vspace*{3mm}
\caption{Examples of downward-bending curtains, as described by problem (\ref{4.1})--(\ref{4.4}), (\ref{4.5}) with $\alpha _{0}=0$. The values of the parameter $\alpha _{0}^{\prime }$ are indicated in the corresponding panels, the values of $v_{0}$ mark the corresponding curves. Panels (a) and (b) correspond to the two horizontal cross-sections of figure \ref{fig6} labelled \textquotedblleft Fig. 5a\textquotedblright\ and \textquotedblleft Fig. 5b\textquotedblright , respectively.}\label{fig5}
\end{figure}

(2) As mentioned before, the trajectories of DB curtains can be computed by
simply choosing a value of $\alpha _{0}^{\prime }$ and shooting the solution
from $l=0$ towards $l\rightarrow \infty $. Numerous examples have been
computed via this approach, some of which are shown in figure \ref{fig5}.

Three features of these examples catch one's eye. First, DB curtains are
wavy (unlike the UB curtains examined above, as well as the DB curtains with
$We\not\approx 1$ examined in B19). Second, curtains with a sufficiently
negative $v_{0}$ become so wavy that they self-intersect (in figure \ref%
{fig5}, only the first intersection is shown; after that, the solution is
meaningless physically). Third, some of the self-intersecting curtains bend
initially upwards (but if their trajectories were extended beyond all
intersections, one would see that they turn downwards eventually).\smallskip

\begin{figure}\vspace*{5mm}
\begin{center}\includegraphics[width=\textwidth]{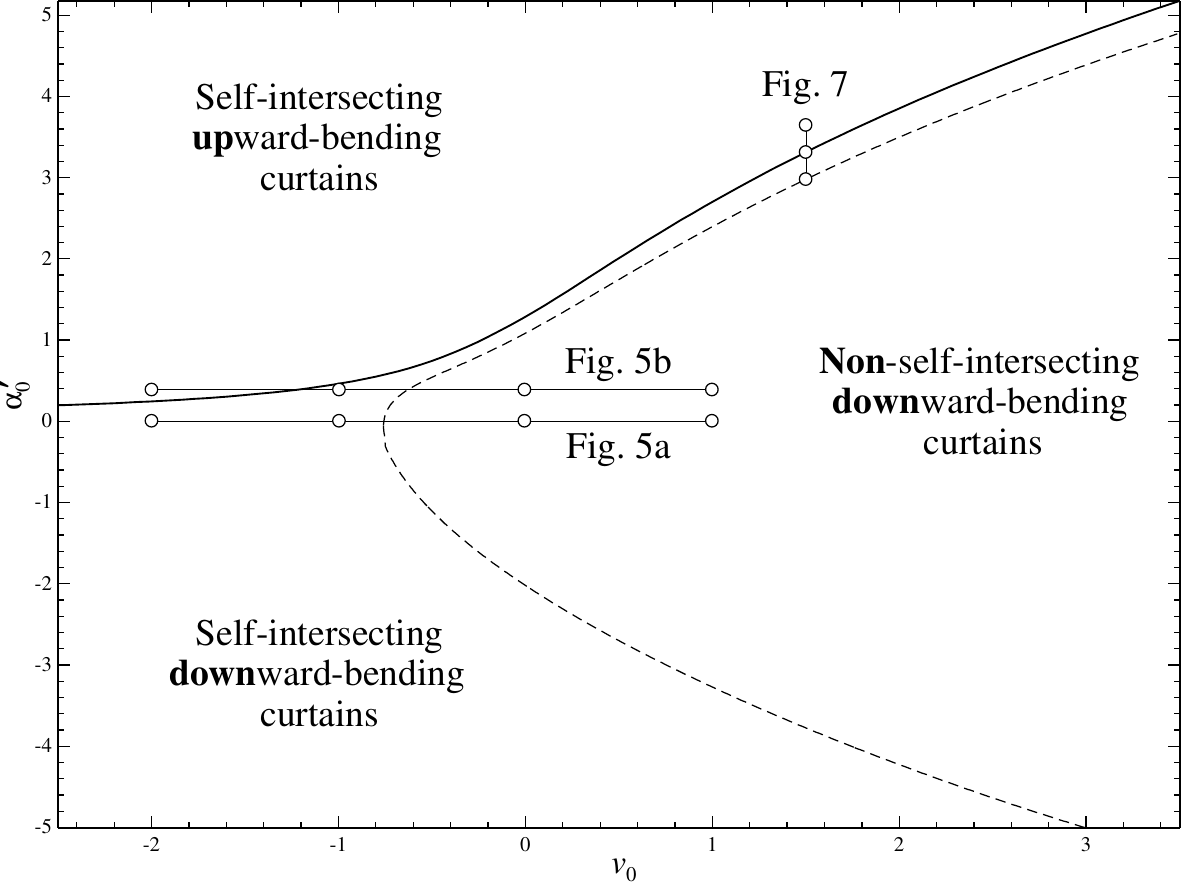}\end{center}\vspace*{3mm}
\caption{Classification of curtains on the $\left( v_{0},\alpha _{0}^{\prime }\right) $ plane, as described by problem (\ref{4.1})--(\ref {4.4}) with $\alpha _{0}=0$. The thick solid curve corresponds to non-self-intersecting upward-bending curtains (it is the same as curve (0) in figure \ref{fig2}). The dashed line bounds the region of self-intersecting curtains. The curtains depicted in figures \ref{fig5} and
\ref{fig7} correspond to the circles connected by thin solid straight lines.}\label{fig6}
\end{figure}

(3) UB and DB curtains can be viewed as different members of the same family
of solutions, as illustrated in figure \ref{fig6}.

From now on, self-intersecting curtains will be classified as UB or DB
depending on how they behave before the first intersection. Then, the curve
corresponding to \emph{non}-self-intersecting UB solutions should be viewed
as a separatrix between \emph{self-intersecting} UB and DB curtains.

\begin{figure}\vspace*{5mm}
\includegraphics[width=\textwidth]{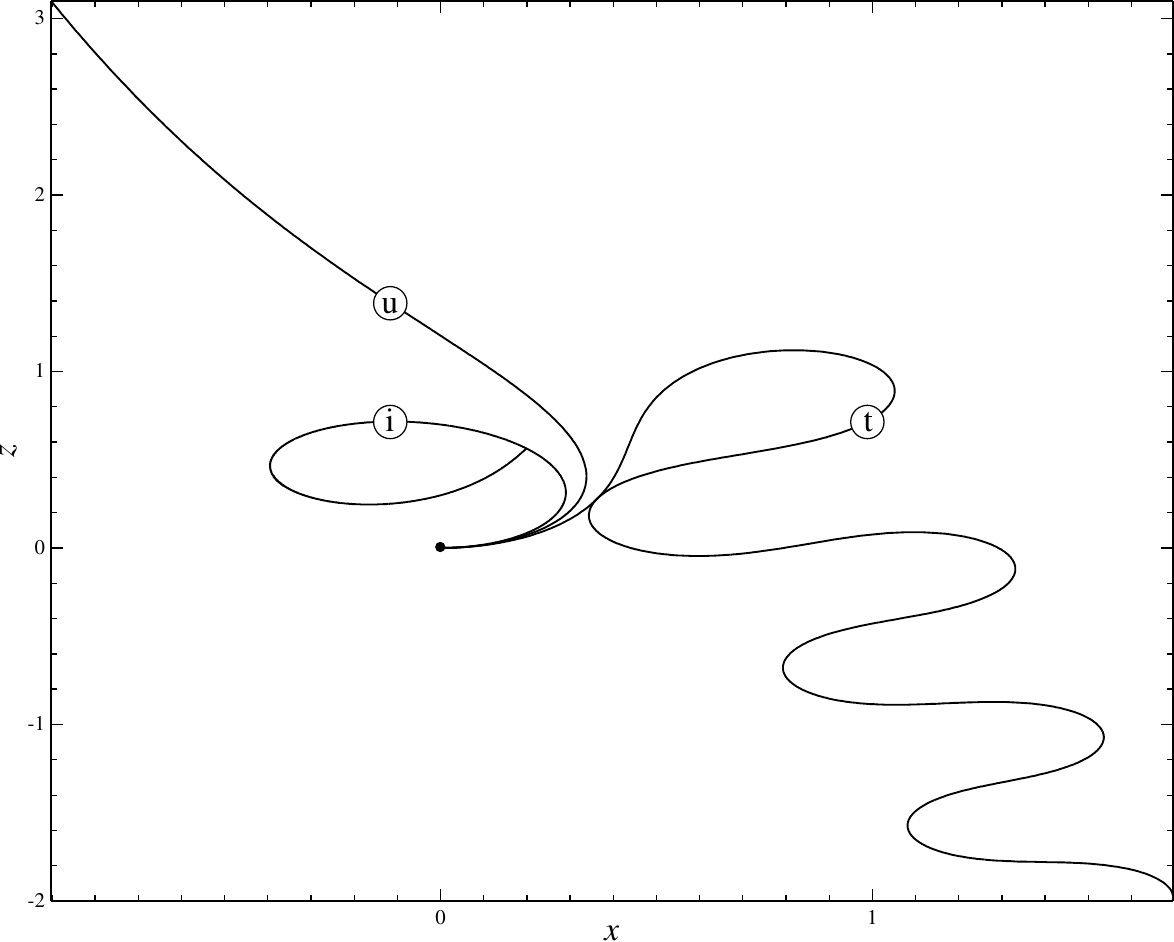}\vspace*{3mm}
\caption{Examples of curtains with $\alpha _{0}=0$ and $v_{0}=1.5$. Curve (u) is the non-intersecting UB curtain ($\alpha _{0}^{\prime }\approx 3.3177$); curve (i) is an example of a self-intersecting UB curtain (with $ \alpha _{0}^{\prime }=3.6467$); curve (t) is the self-touching curtain separating self-intersecting and non-self-intersecting DB curtains ($\alpha _{0}^{\prime }\approx 2.9799$). The solutions depicted correspond to the vertical cross-section of figure \ref{fig6} labelled \textquotedblleft Fig. 7\textquotedblright .}\label{fig7}
\end{figure}

Examples of near-separatrix curtains are shown in figure \ref{fig7}.

\subsection{Comparison with B19\label{section 4.3}}

(1) In B19, UB curtains could only rise to the height where the fluid
velocity vanishes, and so all of the initial reserve of kinetic energy is
used up. In the present model, on the other hand, the leading-order
nondimensional velocity is unity [see (\ref{3.5})] and, thus, cannot vanish
-- so the UB curtains formally rise infinitely high. However, it follows
from (\ref{3.19}) that
\begin{equation*}
u^{(2)}\sim -z\qquad \text{as}\qquad z\rightarrow \infty
\end{equation*}%
-- as a result, once $z$ becomes $\mathcal{O}(\epsilon ^{-2})$, the whole
expansion breaks down. This is the present model's equivalent of the
limiting height of curtains in B19.

The growth of $u^{(2)}$ with growing $z$ invalidates the large-$z$ results
for DB curtains too, but this occurs when they are almost vertical and their
evolution is trivial.\smallskip

(2) One should keep in mind that B19 assumes the characteristic radius of
curvature to be $L=H\,Fr$, whereas the present work considers a shorter
scale, $L=H\,Fr^{1/3}$. In addition, the present work assumes $We\approx 1$.

Thus, the two sets of results should agree only if the limit $We\rightarrow
1 $ is applied to the B19 solutions, whereas the present solutions are
subject to%
\begin{equation}
\alpha _{0}^{\prime }\ll 1,\qquad v_{0}\gg 1.  \label{4.11}
\end{equation}%
These two constraints guarantee that the curtain's nondimensional curvature
is small both near the outlet and globally (as follows from equation (\ref%
{4.1}) with $v_{0}\gg 1$, the solution's global spatial scale is
proportional to $v_{0}$).\smallskip

(3) According to the present results, UB curtains exist for \emph{all} $%
v_{0} $, both positive and negative -- whereas B19 found such solutions only
in the subcritical case $v_{0}<0$.

The apparent contradiction can be resolved if one recalls that, for UB
curtains, $\alpha _{0}^{\prime }$ is a function of $v_{0}$ and $\alpha _{0}$%
, such that $\alpha _{0}^{\prime }\rightarrow \infty $ as $v_{0}\rightarrow
\infty $ (see figure \ref{fig2}). Clearly, such solutions are inconsistent
with limit (\ref{4.11}), and so it comes as no surprise that supercritical
UB curtains have been missed by the asymptotic model of B19.\smallskip

(4) There is another apparent discrepancy between B19 and the present work:
in the former, a single DB curtain was found for given ejection velocity and
angle, whereas the latter found a whole family of solutions, differing from
each other by their values of $\alpha _{0}^{\prime }$.

\begin{figure}\vspace*{5mm}
\includegraphics[width=\textwidth]{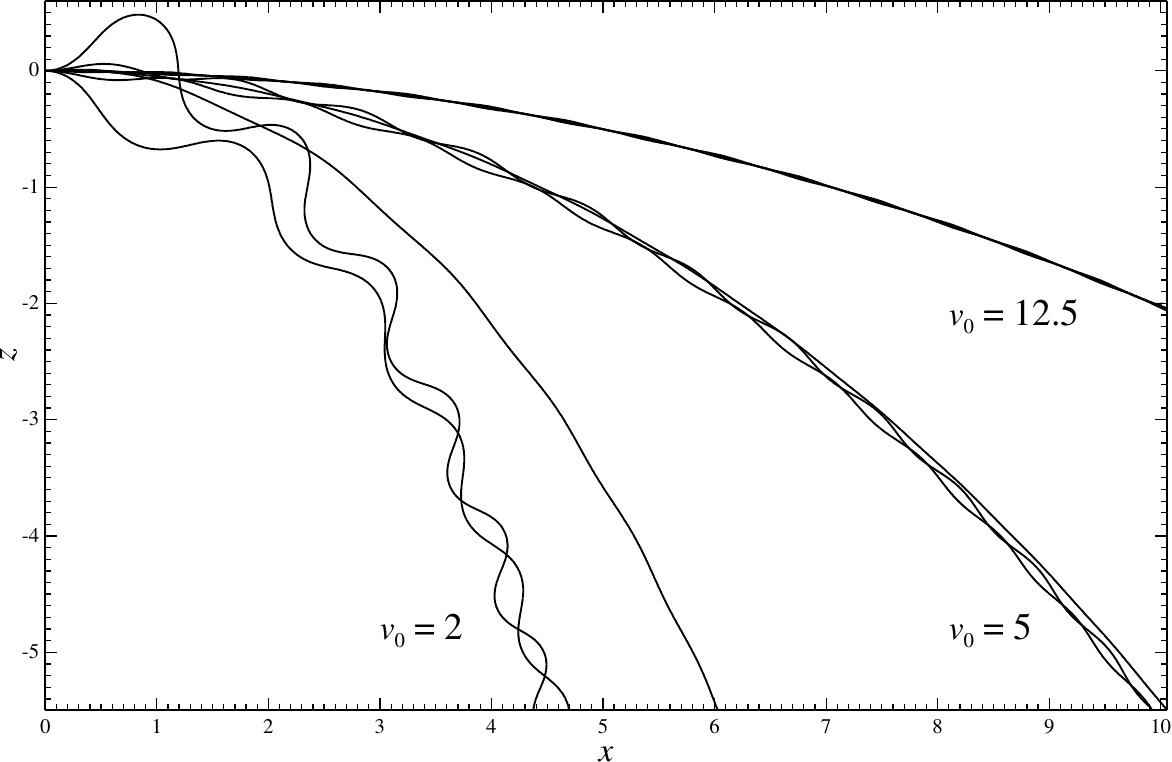}\vspace*{3mm}
\caption{Examples of downwards-bending curtains with $\alpha _{0}=0$. The three curves with $v_{0}=2$ are for $\alpha _{0}^{\prime }=\pm 2.5,~0$, those with $v_{0}=5$ are for $\alpha _{0}^{\prime }=\pm 1,~0$, and those with $v_{0}=12.5$ are for $\alpha _{0}^{\prime }=\pm 0.4,~0$.}\label{fig8}
\end{figure}

To resolve the discrepancy, one should look at the curtains found in the
present work under conditions (\ref{4.11}) -- see figure \ref{fig8}. Observe
that the medium-$v_{0}$ curtains with different $\alpha _{0}^{\prime }$ are
located much closer together than their small-$v_{0}$ counterparts, and the
curtains with the largest value of $v_{0}$ are hardly distinguishable.

Thus, in limit (\ref{4.11}), curtains with different $\alpha _{0}^{\prime }$
collapse onto the same curve, i.e., the dependence on $\alpha _{0}^{\prime }$
becomes weak -- which reconciles the present results with those of B19.

Figure \ref{fig8} also illustrates the fact that limit (\ref{4.11}) makes
curtains less wavy, just as they should be according to B19.

\section{Physical aspects of the solutions found\label{section 5}}

(1) Experiments with capillary curtains are difficult to carry out. The
problem is that, unless the curtain's edges are fixed, they tend to retract
and the curtain contracts into a circular jet.

There are two ways to fix curtains' edges: either by walls %
\citep[e.g.][]{FinnicumWeinsteinRuschak93} or guiding wires %
\citep[e.g.][]{RocheGrandBrunetLebonLimat06,LhuissierBrunetDorbolo16}. Both
worked well with vertical curtains -- but with oblique ones, they probably
would not. Walls imply existence of contact lines which may be pinned and,
thus, can prevent the curtain from assuming its natural shape, whereas wires
certainly prevent the curtain from assuming it. The best option seems to
consist in replacing the curtain with a liquid bell of a large radius (P.
Brunet, private communication) which does not have edges.\smallskip

(2) To come to terms with the counter-intuitive properties of liquid
curtains, one should keep in mind that one's intuition may be misled by
one's everyday experience with jets (those from taps and garden hoses).
There is an important difference between the two types of flows: if the
Weber number of a jet is order-one, it is highly unstable and breaks down
near the outlet by the Plateau--Rayleigh instability, so no-one knows what
shape it would have if it were stable. Capillary curtains, on the other
hand, are presumably stable -- as suggested by experiments with bells %
\citep{BrunetClanetLimat04,JamesonJenkinsButtonSader10} and vertical
curtains \citep{FinnicumWeinsteinRuschak93,RocheGrandBrunetLebonLimat06}, as
well theoretical studies of the latter %
\citep{BenilovBarrosObrien16,GirfoglioDerosaCoppolaLuca17}.\smallskip

(3) The most counter-intuitive feature of DB curtains is the non-uniqueness
of the solution for a given ejection angle and velocity. Each of the
existing solutions represents a sinuous capillary wave, with spatially
dependent amplitude and wavenumber -- such that the wave's local phase
velocity matches that of the flow. These solutions differ from each other by
the wave's amplitude linked to the parameter $\alpha _{0}^{\prime }$: a
larger $\left\vert \alpha _{0}^{\prime }\right\vert $ generally corresponds
to a stronger wave. Setting $\alpha _{0}^{\prime }=0$, however, does \emph{%
not} eliminate the wave entirely (which can be only done by taking the limit
$v_{0}\rightarrow \infty $).

The presence in the solution of a wave component with an arbitrary amplitude
gives rise to two questions: (A) How are these waves generated? (B) Is there
a way to narrow down the family of solutions to the one and only `physically
meaningful' solution?

Question (A) can be answered with certainty: capillary waves are generated
by the variations of the curtain's parameters on its way down -- simply
because a parameter variation in a wave-supporting medium always generates
waves\footnote{%
Interestingly, upward-bending curtains do \emph{not} support waves. This is
evident from the computed examples (see figure \ref{fig3}), as well as the
large-$l$ asymptotics (\ref{4.8})--(\ref{4.10}).}.

As for question (B), the steady state of a curtain probably depends on how
it was created -- or, more generally, on the prior evolution.

One can verify or refute this hypothesis by creating a DB curtain -- waiting
until it becomes steady -- then altering (for a brief period of time) the
ejection angle and/or velocity. Once those have regained their original
values, will the curtain regain its original shape? To a degree of
certainty, this question can be answered through a \emph{gedanken}
experiment -- by solving numerically or asymptotically the full
(evolutionary) version of problem (\ref{3.1})--(\ref{3.4}) with
time-dependent $\alpha _{0}(t)$ and $v_{0}(t)$.\smallskip

(4) The simulation described above may also show how to create a UB curtain
experimentally.

Consider a pair of values $\left( \alpha _{0},v_{0}\right) $ corresponding
to a non-self-intersecting DB curtain -- one of those that do not contradict
our intuition. It corresponds to a point in the right-middle part of figure %
\ref{fig6}, which also shows that the UB\ curtain with the same pair $\left(
\alpha _{0},v_{0}\right) $ has a \emph{larger} near-outlet curvature $\alpha
_{0}^{\prime }$.

Thus, one should be able to create a UB curtain by making a
non-self-intersecting DB curtain increase its $\alpha _{0}^{\prime }$.

This cannot be done directly, as $\alpha _{0}^{\prime }$ is not controlled
in an experiment. One can, however, alter $\alpha _{0}(t)$ and/or $v_{0}(t)$
in such a way that the final steady state has a greater $\alpha _{0}^{\prime
}$ than the initial one.

Indeed, let the injection velocity $v_{0}$ be constant, while the injection
angle $\alpha _{0}(t)$ is slowly increased and then abruptly decreased -- as
one does when cracking a whip. The near-outlet part of the curtain has to
adjust to the change of the injection angle, whereas its main bulk -- due to
its inertia -- will lag behind; as a result, $\alpha _{0}^{\prime }$ should
grow. If it ends up near the value corresponding to the UB curtain, the
resulting steady state should also be close to the UB curtain.

This approach, however, is unlikely to work for strongly subcritical UB
curtains, i.e., those with $v_{0}\ll -1$.

To create one of such, a non-self-intersecting DB curtain is needed to start
from, but such do not exist in this part of the parameter space (see figure %
\ref{fig6}). Computations show that solutions there are highly sensitive to
small variations of the parameters involved, and that a randomly chosen
solution is likely to involve the first self-intersection very close to the
outlet. A strongly subcritical curtain can be guaranteed to have a
reasonably long non-self-intersecting segment only if $\alpha _{0}^{\prime }$
is close to that of the UB curtain with the same $\left( \alpha
_{0},v_{0}\right) $ -- but creating such is as difficult as the UB curtain
itself.\smallskip

(5) The range of physically meaningful solutions (for both DB and UB
curtains ) is likely to narrow if a stability study is carried out.

Still, one should \emph{not} expect that a single solution will emerge as
stable for each $\left( \alpha _{0},v_{0}\right) $. Indeed, stable flows are
generally either nonexistent or occupy a finite-size region in the problem's
parameter space. In the present case, this means that, for some $\left(
\alpha _{0},v_{0}\right) $, there may not be any stable solutions at all --
and for the others, there are infinitely many solutions, corresponding to a
finite-length interval of $\alpha _{0}^{\prime }$.

One can conjecture that unstable curtains are those with high curvature,
where the centrifugal force is too strong to be contained by surface
tension. If this is indeed so, all strongly supercritical ($v_{0}\gg 1$) UB
curtains are unstable (because their near-outlet curvature $\alpha
_{0}^{\prime }$ is large -- see figure \ref{fig2}).

It should also be mentioned that a sufficiently high curvature invalidates
the slender-curtain approximation underlying all of the results
obtained.\smallskip

In view of comments (4)--(5), one should expect UB curtains to be observable
only for moderate $v_{0}$ (or, physically, if $We\approx 1$).

\section{Concluding remarks}

The present work should suffice as a proof concept, but an accurate
description of upward-bending curtains should be based on a more
comprehensive model.

Firstly, \underline{viscosity} should be taken into account. This task
cannot be done in a single step, as two different regimes will have to be
examined. Indeed, let $\nu $ be the liquid's kinematic viscosity. In B19,
its importance was characterised by the nondimensional parameter%
\begin{equation*}
\mu =\frac{\nu u_{0}}{gH^{2}},
\end{equation*}%
and preliminary estimates show that, if%
\begin{equation*}
\mu \epsilon ^{2}\lesssim 1,
\end{equation*}%
viscous forces are comparable to those of inertia and surface tension. In
this case, one should expect the dynamics of curtains with $We\approx 1$ to
be governed by an incremental modification of equation (\ref{3.1}). A
stronger viscosity, however, should dominate the flow and, thus,
significantly change the model.

Secondly, one should extend the present model to \underline{sheared} flows,
as this is the state in which curtains emerge from the outlet. One should
still keep in mind that the transitional region where a sheared (say,
Poiseuille) flow turns into a plug flow can be very small; for jets, for
example, the boundary velocity reaches 50\% of its maximum value after a
distance of only $l\approx 0.04\,H\,Re$
\citep[see][figure
2b]{Goren66,Sevilla11}. If this applies to liquid curtains as well, the
initial shear would have little impact on the global dynamics.

Thirdly, one should examine the \underline{stability} of steady curtains.
Since viscosity is likely to have a significant stabilising effect,
stability study should be carried out only after this effect has been
incorporated into the model.\vspace{0.75cm}

\noindent \textbf{\large Acknowledgments}\medskip

The author is grateful to Andrew Fowler for a comment which turned out to be
crucial for solving this problem.\vspace{0.75cm}

\noindent \textbf{\large Declaration of interests}\medskip

The author reports no conflict of interests.

\appendix

\section{The next-to-leading order of the boundary-layer solution\label%
{Appendix A}}

Observe that, under the condition $\alpha _{0}^{\prime \prime }=0$, the
leading-order solution (\ref{3.38})--(\ref{3.40}) for the cross-stream
velocity amounts to%
\begin{equation}
u_{\tau }^{(0)}=0.  \label{A.1}
\end{equation}%
Substituting this into the boundary-value problem (\ref{3.35})--(\ref{3.37}%
), one can readily deduce expressions for the rest of the leading-order
unknowns,%
\begin{equation}
u_{\tau }^{(0)}=0,\qquad p^{(1)}=-\tau ^{2}\alpha _{0}^{\prime 2},\qquad
\tau _{\pm }^{(2)}=0.  \label{A.2}
\end{equation}%
Next, substitute series (\ref{3.34}) into the full equations (\ref{3.29})--(%
\ref{3.33}), take into account (\ref{A.1})--(\ref{A.2}) and, thus, obtain
the next-to-leading-order equations (the subscript $_{b}$ omitted),%
\begin{equation*}
\dfrac{\partial u_{l}^{(3)}}{\partial l}+\dfrac{\partial p^{(2)}}{\partial l}%
=-\sin \alpha _{0},
\end{equation*}%
\begin{equation*}
\dfrac{\partial u_{\tau }^{(1)}}{\partial l}+2u_{0}\alpha _{0}^{\prime }+%
\dfrac{\tau ^{2}}{2}\left( 8\alpha _{0}^{\prime 3}+\alpha _{0}^{\prime
\prime \prime }\right) +\dfrac{l^{2}}{2}\alpha _{0}^{\prime \prime \prime }+%
\dfrac{\partial p^{(2)}}{\partial \tau }=-\cos \alpha _{0},
\end{equation*}%
\begin{equation*}
\dfrac{\partial }{\partial l}\left( u_{l}^{(3)}+\tau \alpha _{0}^{\prime
}u_{0}+\dfrac{\tau l^{2}}{2}\alpha _{0}^{\prime \prime \prime }\right) +%
\dfrac{\partial u_{\tau }^{(1)}}{\partial \tau }=0,
\end{equation*}%
\begin{equation*}
\dfrac{\partial \tau _{\pm }^{(3)}}{\partial l}-u_{\tau }^{(1)}=0,\qquad
p^{(2)}=\mp \dfrac{\partial ^{2}\tau _{\pm }^{(3)}}{\partial l^{2}}\mp \frac{%
1}{2}\left( 3\alpha _{0}^{\prime 3}+\alpha _{0}^{\prime \prime \prime
}+l^{2}\alpha _{0}^{\prime \prime \prime }\right) \qquad \text{if}\qquad
\tau =\pm 1,
\end{equation*}%
\begin{equation*}
u_{l}^{(3)}=0,\qquad u_{\tau }^{(1)}=0,\qquad \tau _{\pm }^{(3)}=0\qquad
\text{if}\qquad l=0.
\end{equation*}%
One can eliminate from this problem all the unknowns except $\hat{u}_{\tau
}^{(1)}$ and thus obtain%
\begin{equation}
\frac{\partial ^{2}u_{\tau }^{(1)}}{\partial \tau ^{2}}+2l\alpha
_{0}^{\prime \prime \prime }+\frac{\partial ^{2}u_{\tau }^{(1)}}{\partial
l^{2}}=0,  \label{A.3}
\end{equation}%
\begin{equation}
\frac{\partial u_{\tau }^{(1)}}{\partial \tau }-\sin \alpha _{0}=\pm \dfrac{%
\partial ^{2}u_{\tau }^{(1)}}{\partial \tau ^{2}}\qquad \text{if}\qquad \tau
=\pm 1,  \label{A.4}
\end{equation}%
\begin{equation}
u_{\tau }^{(1)}=0\qquad \text{if}\qquad l=0.  \label{A.5}
\end{equation}%
It is convenient to introduce $\hat{u}_{\tau }^{(1)}$ such that%
\begin{equation}
u_{\tau }^{(1)}=\left( 1-\tau ^{2}\right) l\alpha _{0}^{\prime \prime \prime
}+\tau \sin \alpha _{0}+l\frac{\partial \alpha _{0}}{\partial t}+\hat{u}%
_{\tau }^{(1)}.  \label{A.6}
\end{equation}%
In terms of the new unknown, (\ref{A.3})--(\ref{A.5}) become%
\begin{equation}
\frac{\partial ^{2}\hat{u}_{\tau }^{(1)}}{\partial \tau ^{2}}+\frac{\partial
^{2}\hat{u}_{\tau }^{(1)}}{\partial l^{2}}=0,  \label{A.7}
\end{equation}%
\begin{equation}
\frac{\partial \hat{u}_{\tau }^{(1)}}{\partial \tau }=\pm \dfrac{\partial
^{2}\hat{u}_{\tau }^{(1)}}{\partial \tau ^{2}}\qquad \text{if}\qquad \tau
=\pm 1,  \label{A.8}
\end{equation}%
\begin{equation}
\hat{u}_{\tau }^{(1)}=-\tau \sin \alpha _{0}\qquad \text{if}\qquad l=0.
\label{A.9}
\end{equation}%
This boundary-value problem can be solved using the Fourier sine
transformation, but the Fourier transform of $\hat{u}_{\tau }^{(1)}$ would
be singular, and it is not clear how the singularity should be regularised.
Its physical meaning, however, is clear: it describes a semi-infinite wave
generated in the boundary layer and radiated into the outer region. Since
the structure and wavenumber of this wave are easy to find, one can leave
the wave's amplitude arbitrary and `subtract' the wave solution from $\hat{u}%
_{\tau }^{(1)}$. Once the Fourier transform of the modified solution is
found, one can require it to be non-singular and thus find the (so far
undetermined) amplitude.

Following the above plan, introduce $U(l,\tau ,t)$ such that%
\begin{equation}
\hat{u}_{\tau }^{(1)}=U+B^{(1)}\cos k_{\ast }l\,\sinh k_{\ast }\tau
+A^{(1)}\sin k_{\ast }l\,\sinh k_{\ast }\tau ,  \label{A.10}
\end{equation}%
where $k_{\ast }$ is determined by (\ref{3.39}), $B^{(1)}$ is the amplitude
of the wave radiated towards infinity, and $A^{(1)}$ is the amplitude of the
wave coming from infinity -- bouncing off the outlet -- and going back to
infinity. Substituting (\ref{A.10}) into (\ref{A.7})--(\ref{A.9}) and taking
into account (\ref{3.39}) to simplify the boundary condition (\ref{A.8}),
one obtains%
\begin{equation}
\frac{\partial ^{2}U}{\partial \tau ^{2}}+\frac{\partial ^{2}U}{\partial
l^{2}}=0,  \label{A.11}
\end{equation}%
\begin{equation}
\frac{\partial U}{\partial \tau }=\pm \dfrac{\partial ^{2}U}{\partial \tau
^{2}}\qquad \text{if}\qquad \tau =\pm 1,  \label{A.12}
\end{equation}%
\begin{equation}
U=-\tau \sin \alpha _{0}-B^{(1)}\sinh k_{\ast }\tau \qquad \text{if}\qquad
l=0.  \label{A.13}
\end{equation}%
Observe that $A^{(1)}$ remains undetermined, just like its previous-order
counterpart $A^{(0)}$ in expression (\ref{3.38}).

Rewriting (\ref{A.11})--(\ref{A.13}) in terms of the Fourier transform%
\begin{equation*}
\tilde{U}(k,\tau ,t)=\int_{0}^{\infty }U(l,\tau ,t)\sin kl\,\mathrm{d}l,
\end{equation*}%
one obtains%
\begin{equation*}
\frac{\partial ^{2}\tilde{U}}{\partial \tau ^{2}}-k\left( \tau \sin \alpha
_{0}+B^{(1)}\sinh k_{\ast }\tau \right) -k^{2}\tilde{U}=0,
\end{equation*}%
\begin{equation*}
\frac{\partial \tilde{U}}{\partial \tau }=\pm \dfrac{\partial ^{2}U}{%
\partial \tau ^{2}}\qquad \text{if}\qquad \tau =\pm 1.
\end{equation*}%
This boundary-value problem can be readily solved,%
\begin{multline}
\tilde{U}=\frac{\sinh k\tau }{\cosh k-k\sinh k}\left[ \frac{\sin \alpha _{0}%
}{k^{2}}+\frac{B^{(1)}k\left( \cosh k_{\ast }-k\sinh k_{\ast }\right) }{%
k^{2}-k_{\ast }^{2}}\right] \\
-\frac{\tau \sin \alpha _{0}}{k}-\frac{kB^{(1)}\sinh k_{\ast }\tau }{%
k^{2}-k_{\ast }^{2}}.  \label{A.14}
\end{multline}%
Observe that (\ref{A.14}) is not singular at $k=k_{\ast }$ only if%
\begin{equation}
B^{(1)}=\frac{2\sin \alpha _{0}}{k_{\ast }^{2}\sinh k_{\ast }}.  \label{A.15}
\end{equation}%
Together with the expression for the inverse Fourier transform,%
\begin{equation*}
U(l,\tau ,t)=\frac{2}{\pi }\int_{0}^{\infty }\tilde{U}(k,\tau ,t)\sin kl\,%
\mathrm{d}l,
\end{equation*}%
formulae (\ref{A.14})--(\ref{A.15}) complete the solution of problem (\ref%
{A.7})--(\ref{A.9}).

It can be shown that the non-oscillating part of the boundary-layer solution
found matches the outer solution -- but the wave part [the second and third
terms in (\ref{A.10})] implies the inclusion of similar terms in the outer
solution. This has not been done, as such (small and fast-oscillating)
component would affect the global dynamics only when it appears in quadratic
terms -- just as the fast wave component in the Davey--Stewartson system %
\citep{DaveyStewartson74}. In the present problem such terms come up in the
fourth order of the perturbation expansion -- whereas the leading-order
dynamics [equation (\ref{4.1})] emerges in the third order.

Furthermore, if viscosity is introduced, the radiated wave is confined to
the boundary layer and, thus, its effect on the global dynamics is
exponentially weak.

Finally, observe that $u_{\tau }^{(1)}$ is an \emph{odd} function of $\tau $%
, which means that it describes \emph{varicose} capillary waves. Given that $%
u_{\tau }^{(0)}$ is zero, this conclusion applies to the whole
boundary-layer solution.

\section{Numerical solution of problem (\protect\ref{4.1})--(\protect\ref%
{4.4}), (\protect\ref{4.7})\label{Appendix B}}

As argued in \S \ref{section 4.1}, the exponentially growing term in
asymptotics (\ref{4.8})--(\ref{4.10}) has to be eliminated by setting $%
C_{2}=0$, whereas the term involving $\tilde{\alpha}_{3}$ decays faster than
$1/l^{n}$ for any $n>0$. Thus, the large-$l$ asymptotics of the solution of
the full (non-linearised) problem can be sought in the form of a power
series in $1/l$.

Before doing this, however, it is convenient to introduce%
\begin{equation*}
\Delta =z-l,
\end{equation*}%
then define%
\begin{equation*}
\Delta _{\infty }=\lim\limits_{l\rightarrow \infty }\Delta ,
\end{equation*}%
and rewrite (\ref{4.1}) and the second equation of (\ref{4.2}) in the form%
\begin{equation}
\left.
\begin{array}{l}
\dfrac{\mathrm{d}\alpha }{\mathrm{d}l}=\alpha ^{\prime },\medskip \\
\dfrac{\mathrm{d}\alpha ^{\prime }}{\mathrm{d}l}=\alpha ^{\prime \prime
},\medskip \\
\dfrac{\mathrm{d}\alpha ^{\prime \prime }}{\mathrm{d}l}=\alpha ^{\prime }%
\left[ 3\left( \tilde{l}+\Delta -\Delta _{\infty }\right) -\dfrac{1}{2}%
\alpha ^{\prime 2}\right] -3\cos \alpha ,\medskip \\
\dfrac{\mathrm{d}\Delta }{\mathrm{d}l}=\sin \alpha -1,%
\end{array}%
\right\}  \label{B.1}
\end{equation}%
where $\tilde{l}$ is given by (\ref{4.10}). The boundary conditions (\ref%
{4.3}) and the second condition of (\ref{4.4}), in turn, take the form%
\begin{equation}
\alpha (0)=\alpha _{0}\qquad \alpha ^{\prime }(0)=\alpha _{0}^{\prime
},\qquad \alpha ^{\prime \prime }(0)=0\qquad \Delta (0)=0.  \label{B.2}
\end{equation}%
At infinity, let%
\begin{equation}
\left.
\begin{array}{l}
\alpha =\frac{1}{2}\pi +\alpha _{1}\tilde{l}^{-1}+\alpha _{2}\tilde{l}%
^{-2}+\alpha _{3}\tilde{l}^{-3}+\alpha _{4}\tilde{l}^{-4}+\mathcal{O}(\tilde{%
l}^{-5})\medskip \\
\Delta =\Delta _{\infty }+\Delta _{1}\tilde{l}^{-1}+\Delta _{2}\tilde{l}%
^{-2}+\Delta _{3}\tilde{l}^{-3}+\Delta _{4}\tilde{l}^{-4}+\mathcal{O}(\tilde{%
l}^{-5})%
\end{array}%
\right\} \qquad \text{as}\qquad l\rightarrow \infty .  \label{B.3}
\end{equation}%
Substituting asymptotics (\ref{B.3}) into equations (\ref{B.1}), one can
relate the coefficients in (\ref{B.3}) to one of them -- say, $\alpha _{1}$
-- and thus obtain
\begin{equation}
\left.
\begin{array}{c}
\alpha _{2}=0,\qquad \alpha _{3}=-\frac{1}{3}\alpha _{1}^{3},\qquad \alpha
_{4}=\frac{2}{3}\alpha _{1},\medskip \\
\Delta _{1}=\frac{1}{2}\alpha _{1}^{2},\qquad \Delta _{2}=0,\qquad \Delta
_{3}=-\frac{1}{8}\alpha _{1}^{4},\qquad \Delta _{4}=\frac{1}{6}\alpha
_{1}^{2}.%
\end{array}%
\right\}  \label{B.4}
\end{equation}%
Note that asymptotics (\ref{B.3})--(\ref{B.4}) involve three undetermined
parameters: $\Delta _{\infty }$, $\alpha _{1}$, and $\alpha _{0}^{\prime }$
[the last one is `hidden' in $\tilde{l}$, and also appears in equations (\ref%
{B.1}) and boundary conditions (\ref{B.2})].

To solve equations (\ref{B.1}) subject to the boundary conditions (\ref{B.2}%
)--(\ref{B.4}), one should pick a large $l_{\infty }$ and require that three
of the four unknowns $\left( \alpha ,\alpha ^{\prime },\alpha ^{\prime
\prime },\Delta \right) $ coincide at $l=l_{\infty }$ with the values
predicted by the asymptotics (\ref{B.3})--(\ref{B.4}). This way, the
solution will be fixed together with the parameters $\Delta _{\infty }$, $%
\alpha _{1}$, and $\alpha _{0}^{\prime }$.

This approach was realised twice: using the shooting method and the MATLAB
function BVP4c
\citep[based on the three-stage Lobatto IIIa formula --
see][]{KierzenkaShampine01}. The former was found to work only if $l_{\infty
}\lesssim 7$, and so was used only to validate the latter. In either case,
the results were independent of which three of the four unknowns are fixed
at $l=l_{\infty }$.

\bibliographystyle{jfm}
\bibliography{}

\end{document}